\documentclass[10pt,twocolumn,letterpaper]{article}

\usepackage[pagenumbers]{cvpr} %

\newif\ifarxiv
\arxivfalse

\usepackage[dvipsnames]{xcolor}

\newcommand{\tdv}[1]{{#1}}
\newcommand{\del}[1]{}
\newcommand{\maybedel}[1]{}

\def\authorBlock{ 
Radek Daněček \hspace{1cm}
Carolin Schmitt  \hspace{1cm}
Senya Polikovsky  \hspace{1cm}
Michael J. Black \\ 
Max Planck Institute for Intelligent Systems\\
{\tt\small  \{radek.danecek, cschmitt, senya, black\}@tuebingen.mpg.de}
}

\newif\ifreview 
\newif\ifarxiv 
\newif\ifcamera 
\newif\ifrebuttal 

\definecolor{cvprblue}{rgb}{0.21,0.49,0.74}
\usepackage[pagebackref,breaklinks,colorlinks,citecolor=cvprblue]{hyperref}

\usepackage{times}
\usepackage{microtype}
\usepackage{epsfig}
\usepackage{caption}
\usepackage{float}
\usepackage{placeins}
\usepackage{color, colortbl}
\usepackage{stfloats}
\usepackage{enumitem}
\usepackage{tabularx}
\usepackage{xstring}
\usepackage{multirow}
\usepackage{xspace}
\usepackage{url}
\usepackage{subcaption}
\usepackage{xcolor}
\usepackage{afterpage}
\usepackage[hang,flushmargin]{footmisc}

\usepackage{graphicx}
\usepackage{amsmath}
\usepackage{amssymb}
\usepackage{comment}
\usepackage{booktabs} %
\usepackage{numprint}
\usepackage{balance}

\newcommand{\model}{THUNDER\xspace}
\newcommand{\modellong}{\model~(Talking Heads Under Neural Differentiable Elocution Reconstruction)}

\newcommand{\mts}{M2S\xspace}
\newcommand{\svts}{SV2S\xspace}

\newcommand{\todelete}[1]{}

\newcommand{\qheading}[1]{\noindent\textbf{#1}}

\newcommand{\figref}[1]{Fig.~\ref{#1}}
\newcommand{\tabref}[1]{Tab.~\ref{#1}}
\newcommand{\supmat}{Sup.~Mat.\xspace}

\title{Supervising 3D Talking Head Avatars with Analysis-by-Audio-Synthesis}

\author{\authorBlock}

\begin{document}
\newcommand{\shapecoeff}{\boldsymbol{\beta}}
\newcommand{\shapedim}{{\left| \shapecoeff \right|}}
\newcommand{\shapespace}{\mathcal{S}}
\newcommand{\shapespaceexpl}{\mathbb{R}^{\shapedim}}
\newcommand{\posecoeff}{\boldsymbol{\theta}}
\newcommand{\posedim}{{\left| \posecoeff \right|}}
\newcommand{\posespace}{\mathcal{P}}
\newcommand{\posespaceexpl}{\mathbb{R}^{\posedim}}
\newcommand{\jawpose}{\boldsymbol{\theta}_{jaw}}

\newcommand{\jawposerec}{\boldsymbol{\widehat{\theta}}_{jaw}}
\newcommand{\expcoeffrec}{\boldsymbol{\widehat{\psi}}}

\newcommand{\expcoeff}{\boldsymbol{\psi}}
\newcommand{\expdim}{{\left| \expcoeff \right|}}
\newcommand{\expspace}{\mathcal{E}}
\newcommand{\expspaceexpl}{\mathbb{R}^{\expdim}}
\newcommand{\expparam}{\boldsymbol{x}}

\newcommand{\expparamrec}{\boldsymbol{\widehat{x}}}
\newcommand{\expparamnoised}{\boldsymbol{\tilde{x}}}
\newcommand{\verts}{\mathbf{V}}
\newcommand{\vertsrec}{\widehat{\mathbf{V}}}
\newcommand{\vertexcoord}{\verts_m}
\newcommand{\mouthset}{\mathcal{M}}
\newcommand{\upperset}{\mathcal{U}}

\newcommand{\numverts}{n_v}
\newcommand{\numfaces}{n_f}
\newcommand{\template}{\textbf{T}}

\newcommand{\flamev}{M}

\newcommand{\numjoints}{k}
\newcommand{\joints}{\textbf{J}}
\newcommand{\jointregressor}{\mathcal{J}}
\newcommand{\blendweights}{\mathcal{W}}
\newcommand{\blendweightsdim}{\left| \mathcal{W} \right|}

\newcommand{\asrnet}{\mathcal{A}}
\newcommand{\audio}{\mathbf{w}}
\newcommand{\wav}{\audio}
\newcommand{\speechfeat}{\mathbf{a}}

\newcommand{\tmp}{^{1:T}}

\newcommand{\mtsenc}{\mathcal{F}}
\newcommand{\spkenc}{\mathcal{S}}
\newcommand{\spkemb}{\boldsymbol{s}}

\newcommand{\mel}{\boldsymbol{M}}
\newcommand{\speechunit}{\boldsymbol{U}}
\newcommand{\numhubertclasses}{n_{\speechunit}}

\newcommand{\lossmel}{\mathcal{L}_{mel}}
\newcommand{\lossunit}{\mathcal{L}_{unit}}

\newcommand{\lossmts}{\mathcal{L}_{m2s}}
\newcommand{\wmts}{w_{m2s}}
\newcommand{\wmel}{w_{mel}}
\newcommand{\wunit}{w_{unit}}

\newcommand{\denoiser}{\mathcal{D}}
\newcommand{\diffstep}{\mathcal{T}}
\newcommand{\dtrans}{f}

\newcommand{\lossrec}{\mathcal{L}_{rec}}
\newcommand{\lossvel}{\mathcal{L}_{vel}}

\newcommand{\mse}{\textup{MSE}}
\newcommand{\sixd}{\textup{6DOF}}
\newcommand{\vel}{v}

\newcommand{\gauss}{\mathcal{N}(\boldsymbol{\mu}, \boldsymbol{\Sigma})}

\twocolumn[{%
\maketitle
\renewcommand\twocolumn[1][]{#1}
\begin{center}
    \centering
    \vspace{-0.3in}
    \captionsetup{type=figure}
    \includegraphics[trim=0.0cm 2.0cm 1.5cm 2.cm, clip, width=1.99\columnwidth]{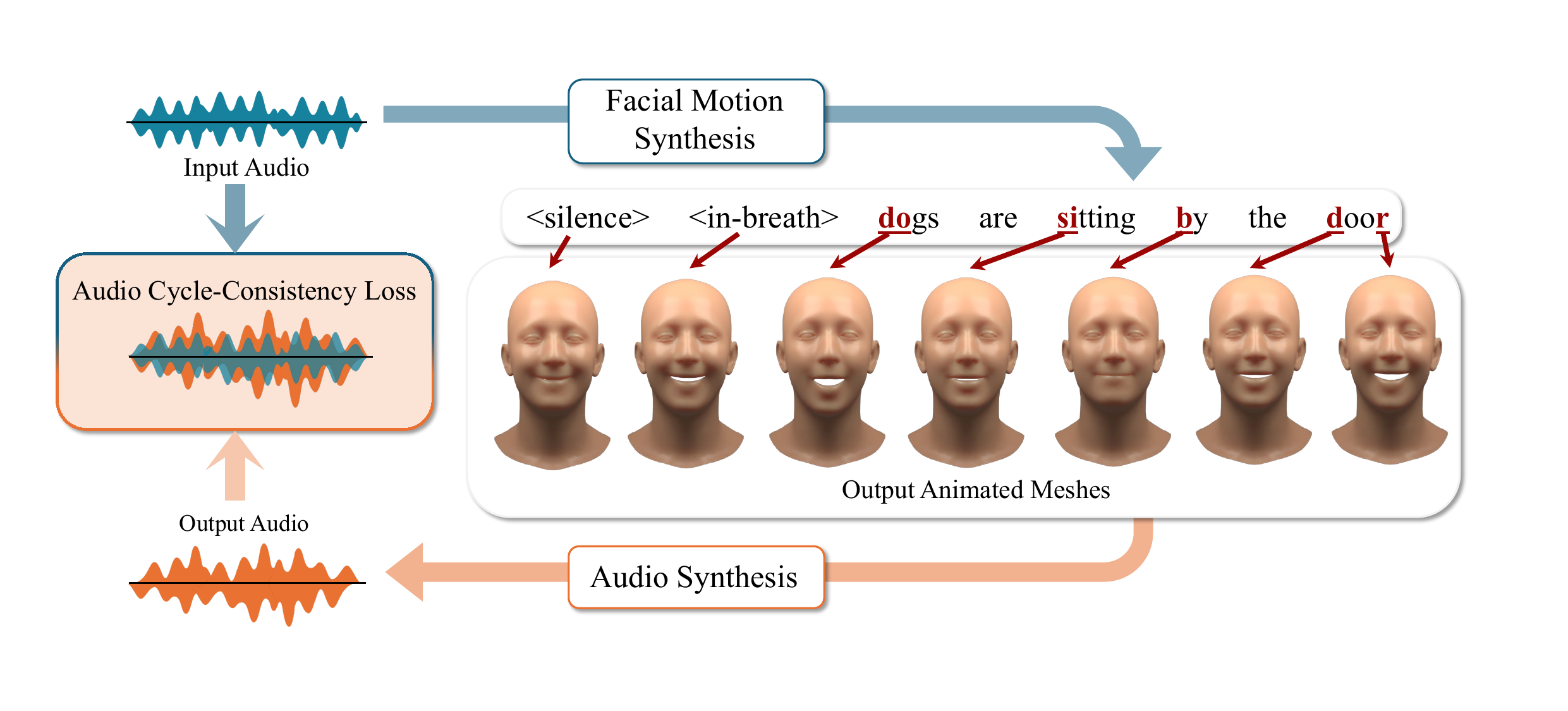}
    \vspace{-0.25cm}
    \caption{
        \model introduces a new paradigm for stochastic generation of 3D talking head avatars from speech with accurate lip articulation and diverse facial expressions.
        Given an input audio, a 3D animation is generated. The animation is then fed into an audio synthesis (mesh-to-speech) model, which generates an output audio representation. The input and output audio representations are compared, creating a novel audio-consistency supervision loop, which we coin as \emph{analysis-by-audio-synthesis}.
    }
    \label{fig:teaser}
\end{center}
}]

\begin{abstract}

In order to be widely applicable, speech-driven 3D head avatars must articulate their lips in accordance with speech, while also conveying the appropriate emotions with dynamically changing facial expressions.  
The key problem is that deterministic models produce high-quality lip-sync but without rich expressions, whereas stochastic models generate diverse expressions but with lower lip-sync quality.
To get the best of both, we seek a stochastic model with accurate lip-sync.
To that end, we develop a new approach based on the following observation: if a method generates realistic 3D lip motions, it should be possible to infer the spoken audio from the lip motion.
The inferred speech should match the original input audio, and erroneous predictions create a novel supervision signal for training 3D talking head avatars with accurate lip-sync. 
To demonstrate this effect, we propose \modellong, a 3D talking head avatar framework that introduces a novel supervision mechanism via differentiable sound production. 
First, we train a novel mesh-to-speech model that regresses audio from facial animation. 
Then, we incorporate this model into a diffusion-based 
talking avatar framework.  
During training, the mesh-to-speech model takes the generated animation and produces a sound that is compared to the input speech, creating a differentiable 
analysis-by-audio-synthesis
supervision loop.
Our extensive qualitative and quantitative 
experiments demonstrate that \model significantly improves the quality of the lip-sync of 
talking head avatars while still allowing for generation of diverse, high-quality, 
expressive
facial animations.

\end{abstract}

\maketitle

\section{Introduction}
\label{sec:intro}

Animating 3D faces from speech has many applications. Examples include the entertainment industry (games and movies), applications in AR/VR/XR (virtual telepresence), e-commerce and perhaps future embodied digital assistants. In order to be widely applicable, the 3D head avatars must, however, appear and act believably and feel alive.
While the methods to animate 3D avatars from speech have come a long way in recent years, there is still a considerable realism gap between real humans and 3D avatars.

Seminal approaches such as VOCA \cite{cudeiro2019capture} or Karras et al.~\cite{karras2017audio}, and others \cite{fan2022faceformer, danecek2023emote, Peng2023_EmoTalk} are deterministic regressors, that predict 3D vertices from the input audio.
Despite achieving good lip-sync, these methods produce facial animations without emotion \cite{cudeiro2019capture, fan2022faceformer} or that are unnaturally static \cite{Peng2023_EmoTalk, danecek2023emote}.
This is in part due to the lack of large-scale and expression-rich training data and in part because a deterministic regressor cannot accurately produce animations of facial motions that only weakly correlate with the audio (such as eyebrow motion, eyeblinks, etc.).
Hence, in order to generate animations that appear lively and believable, a good 3D speech-driven system must model the problem in a non-deterministic manner so that it can capture the many-to-many mapping between speech and facial animation. %

Consequently, recent methods model the problem of 3D speech-driven animation stochastically using relatively small datasets of high-quality 3D scans \cite{richard2021meshtalk, xing2023codetalker, stan2023facediffuser, thambiraja20233diface}; the small training size limits diversity and expressiveness. 
Other methods leverage lower-quality pseudo ground-truth (GT) extracted from larger-scale video datasets \cite{zhao2024media2face, sun2024diffposetalk}.
In pursuing diversity, however, stochastic methods vary the lip motions in ways that may deviate from the audio signal.

The fundamental issue is that lip motions, for the same audio, vary with expression.
We seek a method that generates diverse expressions but is faithful to the audio.
So we ask the question: ``What does it mean to be faithful to the audio?''
This leads us to our key observation; that is, we seek  3D lip motions that {\em actually reproduce the input audio}.
We formalize this notion with a \textbf{new form of supervision} for 3D speech-driven avatars using a novel \textbf{mesh-to-speech} model (\mts); see Fig.~\ref{fig:teaser}. %
Inspired by recent silent-video-to-speech (\svts) methods \cite{choi2023lip2speechunit, kim2023lip, kim2021lipvca}, we propose, to our knowledge, the first \mts model.
Our \mts model regresses the speech audio (or a representation thereof) solely from the animated 3D face. 
In other words, every generated animation now also produces a sound. 
Our rationale is simple: the more similar the produced sound is to the input speech, the more plausible the animation. 
Furthermore, failure to produce the correct sound should provide a good supervision signal.

Our system consists of two stages. 
In the first stage, we train the \mts model to regress audio from facial motion. 
In the second stage, we train a diffusion model to output 3D facial animation from speech. In this stage, we utilize the frozen \mts model. 
The generated facial animation is fed into the \mts model, effectively reproducing the spoken audio. This generated audio (or representation thereof) is then compared to the input audio, creating a self-supervised training loop; this is illustrated in Fig.~\ref{fig:teaser}.
We refer to this as {\em analysis-by-audio-synthesis}.
Our experiments show that leveraging the \mts model results in much better lip-sync quality, while still enabling stochastic production of expressive speech-driven animation.

In summary, our contributions are: 
(1) A novel \textbf{mesh-to-speech} model that regresses sound from 3D facial animations.
(2) A new form of \textbf{cross-modal audio-based self-supervision} via comparison of the representations of the input speech and the animation-produced output speech.
(3) \textbf{\model}, a 3D speech-driven animation method based on diffusion, capable of generating a variety of facial expressions while maintaining accurate lip-sync. 
The code, data annotations and models will be made publicly available 
at \url{https://thunder.is.tue.mpg.de/}.

\section{Related work}

\subsection{Speech-driven 3D animation}

The field of speech-driven 3D facial animation has made significant progress in the last two decades \cite{cao2005expressive, edwards2016jali, edwards2020jalicyberpunk, xu2013practical, taylor2012dynamic, cohen2001animated}.
Here we focus on the recent deep-learning based line of work \cite{pham2017end, pham2017speech, karras2017audio, taylor2017generalized, zhou2018visemenet, cudeiro2019capture, richard2021meshtalk, fan2022faceformer, xing2023codetalker, danecek2023emote, thambiraja20233diface, stan2023facediffuser, zhao2024media2face, sun2024diffposetalk, yang2024probabilistic, Peng2023_EmoTalk, nocentini2024scantalk, aneja2024facetalk, aneja2024gaussianspeech, nocentini2024emovoca, enugi2024enhancing, fan2024unitalker}. 

\qheading{Deterministic neural methods.}
Karras et al.~\cite{karras2017audio} were the first to utilize deep learning by training a temporal convolutional network (TCN) to predict 3D face vertices.
VOCA \cite{cudeiro2019capture} employs a pre-trained automatic speech recognition (ASR) network to regress face vertex offsets from audio, achieving good lip-sync for multiple speaking styles. 
Many follow-up methods use a similar determinstic paradigm \cite{fan2022faceformer, bala2022imitator, FaceXHuBERT_Haque_ICMI23, Peng2023_EmoTalk, danecek2023emote, peng2023selftalk, nocentini2024scantalk, nocentini2024emovoca, xu2024kmtalk, fu2024mimic, aneja2024gaussianspeech, enugi2024enhancing, fan2024unitalker, sungbin2024multitalk, kim2025MemoryTalker}.

\qheading{Stochastic neural methods.}
The first model to approach the problem stochastically was MeshTalk \cite{richard2021meshtalk}. MeshTalk makes use of a pretrained discretized facial motion prior along with an explicit supervision mechanism that supports generation of motions that only have a weak correlation with the audio (eyebrow motion, etc.). 
\del{CodeTalker \cite{xing2023codetalker} upgrades the FaceFormer architecture by predicting codebook classes of a pretrained VQ-VAE. Both CodeTalker and }
MeshTalk employs autoregressive predicition and one can therefore sample the distribution of the next token to generate variety. 

Thanks to the tremendous success of Diffusion Models in the image domain \cite{ho2020denoising, rombach2022stable} they are now widely used for 3D human body animation \cite{tevet2023hmd, chen2023mld}, as well as speech-driven animation \cite{stan2023facediffuser, thambiraja20233diface, sun2024diffposetalk, zhao2024media2face}.  
The diffusion methods also employ transformer-based audio feature extractors \cite{baevski2020wav2vec, hsu2021hubert} to condition the denoiser. 
FaceDiffuser \cite{stan2023facediffuser} employs a GRU-based denoiser, 3DiFACE \cite{thambiraja20233diface} uses a TCN-based denoiser,
and DiffPoseTalk \cite{sun2024diffposetalk}, Media2Face \cite{zhao2024media2face} and FaceTalk \cite{aneja2024facetalk} employ a transformer decoder architecture \cite{vaswani2017attention} passing the audio condition to the denoiser via cross-attention.  
\tdv{Media2Face and DiffPoseTalk are the most similar to \model --- they use pseudo-GT on a large scale and utilize a transformer-decoder to denoise the animation in the 3DMM space.}

\qheading{Controlling the output animation.}
The task of controlling or editing the output animation is of great utility in production as manually editing animations is a laborious process. 
Many methods use a simple one-hot encoding of training subjects, which is mapped to a ``style embedding'' with a learnable layer. This embedding conditions the decoder to match the speaking style of the corresponding training subject \cite{cudeiro2019capture, fan2022faceformer, xing2023codetalker, bala2022imitator}. This paradigm has been extended to control the emotion and emotion intensity \cite{danecek2023emote, Peng2023_EmoTalk}. 
DiffPoseTalk \cite{sun2024diffposetalk} uses contrastive learning to produce a style vector from a reference animation, lifting the subject-ID conditioning limitation.
Media2Face \cite{zhao2024media2face} employs CLIP features \cite{Radford2021clip}, which can be extracted from an image, or text prompts to condition the denoising process. 
3DiFACE \cite{thambiraja20233diface} demonstrates that the output of a diffusion model can be guided by a sparse set of keyframes. 
In this paper, instead of focusing on controlling the animation, we present a general method to improve lip-sync quality, which can be used in conjunction with the above control mechanisms.

\qheading{Lip animation experts.}
\tdv{A few methods attempt to design specific expert mechanisms to improve the lip-sync quality. 
EMOTE \cite{danecek2023emote} uses an image-based lip-reading network on a differentiably-rendered output image in there, but the need for differentiable rendering of videos in-the-loop makes the approach slow and GPU-memory intensive.
SelfTalk \cite{peng2023selftalk} proposes augmenting the FaceFormer architecture with a language decoder and language-based losses. 
A recent work by Chae et al.~\cite{chae2025perceptually} employs masked auto-encoding and contrastive learning (akin to CLIP) to train a speech-mesh representation. 
The authors showed that the network can improve lip-animations in deterministic systems like FaceFormer or SelfTalk.
To the best of our knowledge, no existing method attempts to synthesize audio for cycle-consistency, or employ lip animation experts in stochastic speech-driven animation.
}

\qheading{Datasets.}
Most methods \cite{karras2017audio, cudeiro2019capture, fan2022faceformer, 
xing2023codetalker, bala2022imitator} have made use of high quality 3D scans synchronized with audio, such VOCASET \cite{cudeiro2019capture}\del{, BIWI \cite{eth_biwi_00760}} or Multiface \cite{wuu2022multiface} \tdv{or Kinect-captured BIWI \cite{eth_biwi_00760}}.
\del{While these datasets are %
of high quality, they}
\tdv{These datasets} are expensive to acquire and hence are limited in number of subjects, and richness of facial expressions. 
\maybedel{
The dataset design is often a limiting factor, as the samples are usually 
sentences taken out of context and hence the subjects do not express themselves in the most natural ways. 
}
As a consequence, even the stochastic methods that leverage these datasets 
\cite{xing2023codetalker, stan2023facediffuser}, 
suffer from lack of natural diversity of expressions. 
Thanks to the significant improvement of fast face reconstruction regressors \cite{EMOCA:CVPR:2021, MICA:ECCV2022, filntisis2022visual, zhang2023tokenface, Retsinas_2024_CVPR_Smirk},
recent methods have turned to using pseudo-ground-truth (pGT) reconstructions from videos \cite{danecek2023emote, Peng2023_EmoTalk, sun2024diffposetalk, zhao2024media2face}.
While the pGT does not reach the quality of 3D scans, 
\tdv{its quality is now sufficient.}
Video datasets are numerous, allowing more data to be acquired, which benefits data-hungry stochastic models.

\subsection{Silent-video-to-speech (SVTS)}
To the best of our knowledge, the prediction of voice from a sequence of 3D facial shapes has never been explored. 
However, recent years have seen tremendous progress on the task of speech audio prediction from silent videos. 
Early methods \cite{cornu2015speechfromfeaturs, Oneata2021speaker, hong2021speechrec, kim2021lipvca, Michelsanti2020vocoderbased, Yadav2021speech, Vougioukas2019videodriven, um2023facetron} focus on small in-the-lab datasets with predefined scripts and limited number of speakers (GRID \cite{cooke2016grid}, TCD-TIMIT \cite{Harte2015tcdtimit}) or script-unconstrained but single-speaker models \cite{prajwal2020cvrplearning}.

SVTS \cite{carrillo2022svts} was the first method to produce intelligible audio on large-scale in-the-wild datasets such as LRS2 and LRS3 \cite{afouras2018lrs3} by regressing spectrograms from mouth crops. The spectrograms are converted to the final waveform with a pretrained vocoder\cite{choi2023lip2speechunit}. 
Follow-up methods improve the prediction quality \cite{kim2023lip} by adding a surrogate ASR loss or predicting features from an ASR model \cite{choi2023lip2speechunit}
The most recent architectures are based on diffusion, producing audio 
often
indistinguishable from real speech \cite{choi2023diffv2s, yemini2023lipvoicer}. These, however, are not suitable for a self-supervised reconstruction loop, due to the iterative nature of diffusion models.
Hence, thanks to its simple, yet efficient feed-forward design and a relatively straightforward architecture and training, we base our mesh-to-speech model on Choi et al.\cite{choi2023lip2speechunit}. 

\subsection{Audio cycle consistency}
Cycle consistency losses on audio have been applied before, for tasks like voice conversion \cite{Kaneko2020CycleGAN-VC, Kaneko2020CycleGAN-VC2, Kaneko2020CycleGAN-VC3} and 
disentangled representation learning \cite{ji2021evp, Chhatre2024amuse}. 
Many different self-supervised audio learning methods exist (see survey \cite{liu2022audioself}).
However, to our knowledge, we are the first to propose cross-modal cycle consistency between audio and 3D facial motion.

\section{Preliminaries}

\subsection{Face model}
\model uses the FLAME face model \cite{FLAME:SiggraphAsia2017}, which is a statistical 3D morphable model (3DMM). 
FLAME provides a compact representation of facial shapes and expressions and is defined as a function:
\begin{equation}
    \flamev(\shapecoeff, \posecoeff, \expcoeff)\rightarrow (\verts, \mathbf{F}),
\end{equation}
where the inputs are shape coefficients $\shapecoeff \in \mathbb{R}^\shapedim$, expression coefficients $\expcoeff \in \mathbb{R}^\expdim$ and rotation vectors for $\numjoints=4$ joints $\posecoeff \in \mathbb{R}^{3\numjoints+3}$. 
FLAME outputs a 3D mesh with vertices $\verts \in \mathbb{R}^{\numverts \times 3}$ and triangles $ \mathbf{F} \in \mathbb{R}^{\numfaces \times 3}$.
Since we do not focus on head movement, we only use the jaw joint rotation $\jawpose$. 
For brevity, we refer to the expression coefficients and jaw rotation as expression parameters $\expparam = [\expcoeff | \jawpose]$.

\subsection{Speech feature extraction}
Following previous work on speech-driven facial animation \cite{fan2022faceformer, xing2023codetalker, bala2022imitator, danecek2023emote, zhao2024media2face} we use Wav2Vec2.0 \cite{baevski2020wav2vec} as our audio feature extractor. 
It consists of a temporal convolution network (TCN) that extracts audio features at 50Hz and a transformer encoder that processes these features. 
Similarly to previous methods, we resample the TCN feature to 25Hz and feed it to the transformer to obtain the final speech feature. 
Formally, we define this as: 
$
     \asrnet(\audio) \rightarrow \speechfeat \tmp,
$
where $\asrnet$ denotes the Wav2Vec 2.0 network, $\audio$ is the input waveform, and 
$ \speechfeat \tmp \in \mathbb{R}^{T \times d_s}$, with $T$ denoting the number of frames at 25Hz and $ d_s=768 $ is the feature dimension.

\subsection{Speaker Embedding} 
A speaker embedding is a numerical representation of a speaker's unique vocal characteristics encoded into a vector. 
This vector captures features such as tone, pitch, speaking style, and other vocal attributes, enabling the differentiation and recognition of individual speakers. 
Among other applications, speaker embeddings are used in the latest \svts models \cite{carrillo2022svts, kim2023lip, choi2023lip2speechunit}.
Following Choi et al.~\cite{choi2023lip2speechunit}, we leverage an off-the-shelf speaker embedding extractor \cite{RealTimeVoiceCloning}. 
We define the extractor as: 
$
    \spkenc(\wav) \rightarrow \spkemb,
$
with $\spkenc$ %
and $\spkemb$ denoting 
embedding extractor
and 
the 
embedding vector, respectively. 

\subsection{Speech Units} 
The term speech units refers to discrete linguistic units identified through a self-supervised learning process. 
Like Choi et al.~\cite{choi2023lip2speechunit}, we leverage speech units predicted by a pretrained HuBERT \cite{hsu2021hubert} model and its associated feature clustering model. 
First, features are extracted and subsequently clustered into $\numhubertclasses$ units. 
We define speech units as one-hot class vectors 
$\speechunit \in [c_1, c_2, \dots, c_{\numhubertclasses}]$ 
where only one of $c_i, i=1 \dots c_{\numhubertclasses}$ is $1$ and the rest are $0$. 
Speech units are predicted at 50Hz.

\section{Method}

\begin{figure}
    \centering
    \includegraphics[trim=24 0 0 0, clip, width=1.\columnwidth]{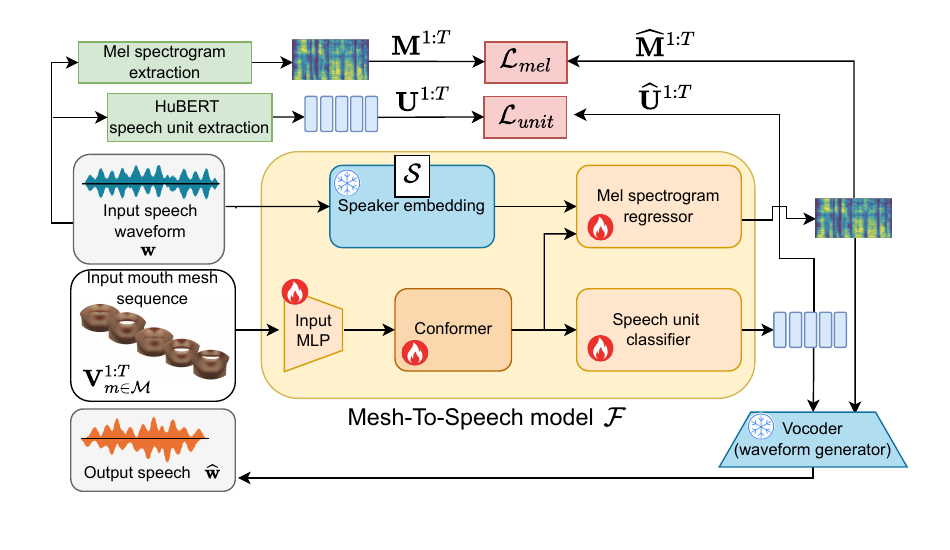} 
    \vspace{-0.4in}
    \caption{
        \textbf{Mesh-to-speech architecture.} It takes a sequence of mouth shapes as input, along with a speaker embedding feature, to produce the output speech units and spectrograms. These are used to compute a loss (top) and to produce the reconstructed audio (bottom) using a pretrained vocoder. 
    }
    \label{fig:mts_arch}
\end{figure}

\model is trained in two stages. 
First, a 
mesh-to-speech (M2S)
network is trained to regress audio from facial motion. 
Second, a diffusion model is trained to output 3D facial animation. %
The output animation is then fed to the frozen \mts model producing output audio representations.
This allows us to design a loss between the input and output representations, which we utilize in training the talking head system.

\subsection{Mesh-To-Speech}
Regressing audio from 3D facial motion has, to our knowledge, 
never been done. 
To narrow the design space, we draw inspiration from
the recent \svts model by Choi et al.~\cite{choi2023lip2speechunit}.

\qheading{Architecture.}
The architecture consists of an input feature encoder, a conformer sequence encoder and two prediction heads, a speech unit classifier, and a mel spectrogram regressor. 
These two outputs contain enough information that a pretrained off-the-shelf vocoder \cite{choi2023lip2speechunit} can turn them into output audio. 
Since the input is not a video, but a 3D animation, we replace the original lip video ResNet encoder that Choi et al.~\cite{choi2023lip2speechunit} use with an MLP that takes 3D lip vertex coordinates as the input. 
We keep the rest of the architecture the same. 
The \mts architecture is depicted in \figref{fig:mts_arch} and can be written
\vspace{-0.3cm}
\begin{equation}
    \mtsenc(\verts_{m \in \mouthset} \tmp, \spkemb ) \rightarrow (\widehat{\mel}\tmp, \widehat{\speechunit}\tmp), 
\end{equation}
where $\mtsenc(.)$ denotes the \mts network, $\mouthset$ is a subset of mouth vertices. $\widehat{\mel}\tmp ,\widehat{\speechunit}\tmp$ denote the output sequence of mel spectrograms and speech units. 
Finally, $\widehat{\mel}\tmp$ and $\widehat{\speechunit}\tmp$ can be passed to an off-the-shelf vocoder to generate the output waveform $\widehat{\wav}$.

\qheading{Supervision.}
We follow the same supervision scheme as Choi et al.~\cite{choi2023lip2speechunit}. 
The loss consists of two terms. The first term is an L1 loss between the input and output mel specrograms:
\vspace{-0.3cm}
\begin{equation}
    \lossmel = \| \mel\tmp - \widehat{\mel}\tmp \|_1 .
\end{equation}
The second term is cross entropy between the predicted speech unit classification vectors and the GT speech units:
\vspace{-0.3cm}
\begin{equation}
    \lossunit = \frac{1}{T}\sum_{t=1}^{T} \sum_{c=1}^{\numhubertclasses} \speechunit_c^t \ \textup{log} \ p (\widehat{\speechunit}_{c}^{t}),
\end{equation}
\vspace{-0.1cm}
with weights $\wmel=10$ and $\wunit=1$. The final loss is: %
\vspace{-0.1cm}
\begin{equation}
\label{eq:lossmts}
    \lossmts = \wmel \lossmel + \wunit \lossunit .
\end{equation}

\subsection{Speech-Driven Facial Animation Diffusion} 

\begin{figure*}
    \centerline{
    \includegraphics[trim=0.0cm 0.0cm 1.5cm 0cm, clip, width=2.1\columnwidth]{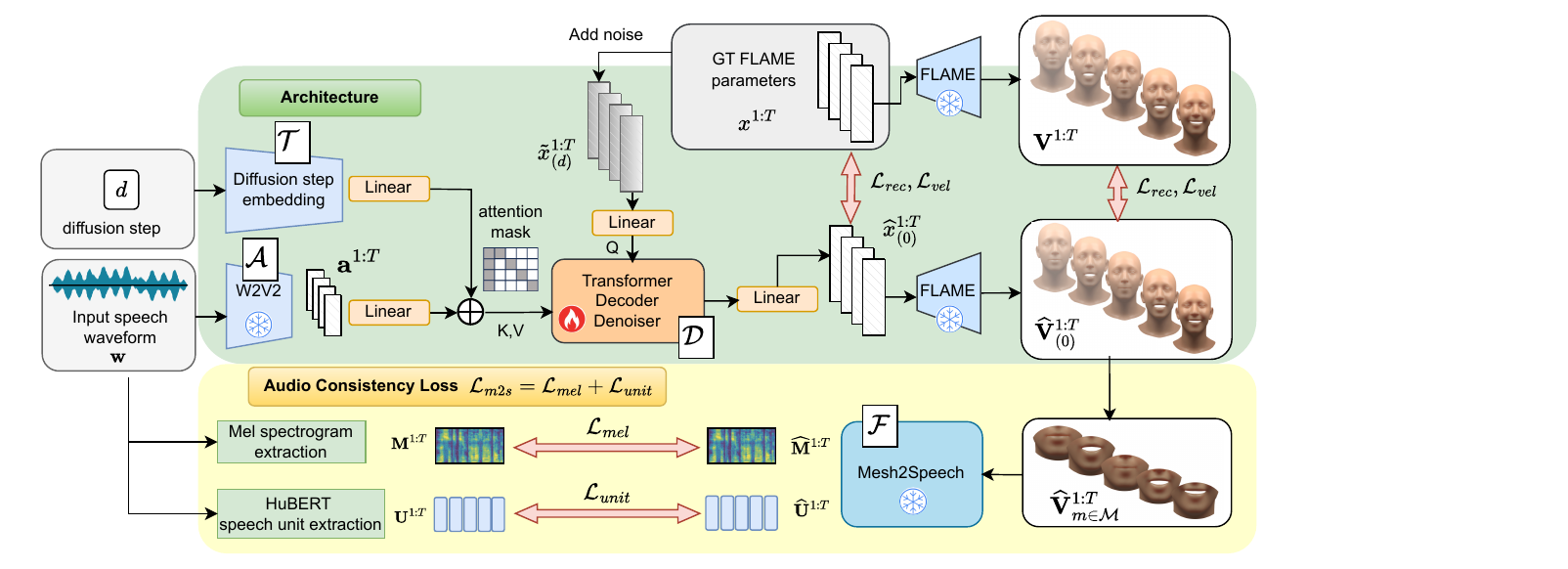}}
    \vspace{-0.1in}
    \caption{
        \textbf{\model architecture.} The upper part of the figure (green) depicts the architecture of the diffusion model and the lower part (yellow) illustrates the application of the audio consistency loss. Gray boxes indicate the input to the system. Trainable components are highlighted in orange and frozen ones in blue. 
    }
    \label{fig:diffusion_arch}
\end{figure*}

\qheading{Architecture.}
The core of \model consists of a denoiser in the form of a transformer decoder
\cite{vaswani2017attention}. 
Similar to other contemporary 3D animation methods (e.g.~Zhao et al.~\cite{zhao2024media2face} and Sun et al.~\cite{sun2024diffposetalk}), \model processes a noisy sequence of expression parameters $\expparamnoised_{(d)}\tmp$, where $d$ represents the diffusion timestep. 
These parameters serve as the \textit{query} inputs to the transformer.
The \textit{keys} and \textit{values} for the transformer are comprised of the input audio features $\speechfeat\tmp$ and a diffusion step embedding $\diffstep(d)$, concatenated along the temporal dimension. 
The function $\diffstep$ is defined as a sinusoidal timestep function similar to Ho et al.~\cite{ho2020denoising}. 
We do not incorporate additional animation-controlling conditions (such as style vectors \cite{cudeiro2019capture, fan2022faceformer, xing2023codetalker}, emotion vectors \cite{danecek2023emote, Peng2023_EmoTalk} or CLIP embeddings \cite{zhao2024media2face}) as this is not the goal of this work. 
However, the design can easily be augmented with additional input conditions. 
All inputs are projected into the transformer's feature dimension, $\dtrans=128$, using dedicated learnable affine layers. 
Positional encoding within the transformer's self-attention utilizes the ALiBi mechanism \cite{press2021train}. 
The cross-attention layers use a diagonal binary mask to facilitate the attention mechanism, and no additional positional encoding.
The model predicts the reconstructed expression parameters $\expparamrec_{(0)}\tmp$. 
The transformation can be formally described by the following equation:
\vspace{-0.1cm}
\begin{equation}
    \denoiser(\expparamnoised_{(d)}\tmp, d, \speechfeat\tmp) \rightarrow \expparamrec_{(0)}\tmp .
\end{equation}
Here, $d \in \{1, \dots, D\}$ indicates the diffusion timestep, and $\expparamrec_{(0)}\tmp$ is the denoised prediction. Unlike some methods that predict noise, this model directly predicts the fully denoised space, which is used to compute the 3D mesh sequence:
$
    \flamev(\shapecoeff=\mathbf{0}, \expparamrec_{(0)}\tmp) \rightarrow \vertsrec_{(0)}\tmp
$.
This approach allows the model to define losses directly on the final 3D mesh space, crucial for leveraging \mts. 
See \figref{fig:diffusion_arch} for an overview.

\qheading{Training.}
During training, a diffusion step $d$ is sampled and $\expparamnoised_{(d)}\tmp$ is computed from the clean expression parameters $\expparam\tmp$ using a noise schedule that blends the original data with Gaussian noise based on predefined $\alpha$ values: %
\begin{equation}
\expparamnoised_{(d)}\tmp = \sqrt{\alpha_d} \expparam\tmp + \sqrt{1 - \alpha_d} \cdot \epsilon, \ \epsilon \sim \gauss),
\label{eq:noising}
\end{equation}
where $\alpha_d$ represents the proportion of the original signal's variance preserved at diffusion timestep $d$, and $\epsilon$ is drawn from a Gaussian distribution with predefined $\boldsymbol{\mu}$ and $\boldsymbol{\Sigma}$. 
This approach incrementally corrupts the clean parameters, enabling the model to learn to denoise varying levels of noise.
The model is supervised with several loss terms - reconstruction and velocity losses in expression space and 3D vertex space, and also the \mts loss (Eq. \eqref{eq:lossmts}).
The reconstruction loss is given as: 
\vspace{-0.2cm}
\begin{equation*}    
    \lossrec = \ \|\verts\tmp-\vertsrec_{(0)}\tmp\|_2^2 + \|\expcoeff\tmp-\expcoeffrec_{(0)}\tmp\|_2^2+ 
     \ \| \jawpose\tmp-{\jawposerec\tmp}_{(0)} \|_2^2 .
\end{equation*}

The velocity loss is also an MSE error but computed on the velocities of the individual components:
\vspace{-0.2cm}
\begin{align*}    
    \lossvel = & \ \|\vel(\verts\tmp) - \vel(\vertsrec_{(0)}\tmp)\|_2^2 + \|\vel(\expcoeff\tmp) - \vel(\expcoeffrec_{(0)}\tmp)\|_2^2+ \\
    & \ \| \vel(\jawpose\tmp) - \vel({\jawposerec\tmp}_{(0)} )\|_2^2,
\end{align*}
where $\vel$ computes the velocity of any input sequence $\mathbf{u}\tmp$ $\vel(\mathbf{u}\tmp) = \mathbf{u}^{2:T} - \mathbf{u}^{1:T-1}$.
The final loss is given as: 
\begin{equation}
    \mathcal{L}_{total} = \wmts \lossmts + \lossrec  + \lossvel, 
\end{equation}
with $\wmts=1$. The model is trained using classifier-free guidance \cite{ho2022cfg}. The audio condition is randomly dropped and replaced with a learnable vector with 20\% probability.

\qheading{Inference.} At inference time, we initialize the first step by sampling the Gaussian noise: $\expparamrec_{(D)} \sim  \gauss $. 
We then proceed to denoise for $D$ diffusion steps. 
In each step, we first employ the denoiser $ \denoiser(\expparamnoised_{(d)}\tmp, d, \speechfeat\tmp) $ to obtain $ \expparamrec_{(0)}\tmp$. 
Then, we compute $\expparamnoised_{(d-1)}$ by adding the expected amount of noise back to obtain $\expparamnoised_{(d-1)}$.
Since \model is trained with classifier-free guidance, it is possible to trade-off fidelity of the lip-sync for increased diversity by combining $ s_{a} \denoiser(\expparamnoised_{(d)}\tmp, d, \speechfeat\tmp) + (1-s_a) \denoiser(\expparamnoised_{(d)}\tmp, d, \emptyset) $ during the denoising process. 
However, we find our results are diverse enough without this and so we effectively set $s_a=1$.

\section{Experiments}

\subsection{Datasets}
The selection of datasets for \model is challenging. 
\del{On the one hand, }4D scan datasets such as VOCASET \cite{cudeiro2019capture} provide good GT but are too small to train generative models.
\del{On the other hand, i}
In order to have a high-enough quality geometry from which speech can be predicted, 
\del{we cannot rely on in-the-wild datasets, as reconstruction artifacts decrease the quality of the GT and hence our \mts and talking head models. Hence, }
we use a collection of public in-the-lab captured video datasets commonly used for lip-reading, namely 
TCD-TIMIT \cite{Harte2015tcdtimit}, 
RAVDESS \cite{livingstone_2018_1188976} 
and GRID \cite{cooke2016grid}, 
totaling 117 subjects,  %
uttering 42140
short sentences, 
totaling 3535816 frames at 25FPS (over 39 hours).
These datasets provide clear imagery without in-the-wild complexity and can be reconstructed by off-the-shelf methods to a satisfactory degree. 
Following EMOTE \cite{danecek2023emote}, we employ the publicly available INFERNO framework \cite{inferno2023}, which contains a SOTA face reconstruction system \cite{MICA:ECCV2022, EMOCA:CVPR:2021, filntisis2022visual}.
\tdv{We include an expanded discussion on face tracker selection in Sec.~\ref{sec:data_acquisition} in \supmat}
We split each of the datasets by subjects, 70\% for training and 15\% each for validation and testing. 
We refer to this data as THUNDERSET.
\tdv{Additionally, we also train and test \mts and \model on a more challenging TFHP \cite{sun2024diffposetalk}, which contains sequences in-the-wild and head pose.
}

\subsection{Mesh-To-Speech}

\qheading{Implementation details.}
We initialize the conformer and prediction heads with pretrained weights from Choi et al.~\cite{choi2023lip2speechunit}, while the input MLP is trained from scratch. 
We train the model with the Adam optimizer \cite{Kingma2015} for 20 epochs and select the checkpoint with the lowest validation loss (usually reached in 5 epochs). The training batch size is 16 and the input sequence is trimmed to 125 frames at 25FPS.

\qheading{Quantitative comparison.} 
Since the \mts model is the first of its kind, we compare \mts to a recent \svts model \cite{choi2023lip2speechunit} which we finetune on the videos from our training dataset for fair comparison.
Note that FLAME does \emph{not} model the teeth or tongue, which makes the \mts task arguably harder than 
\svts.
Teeth and tongue, which may be visible on video, are critical for production of certain phonemes, such as labiodentals (/f/, /v/), 
or alveolar phonemes (/s/, /z/, /t/, /d/, /l/) \cite{Ladefoged2011phonetics}.
Consequently, we do not expect \mts to match the performance of 
\svts methods.

Table 
\ref{tab:mesh2speech_ablation_public} 
reports the standard video-to-speech metrics, namely Short-Time Objective Intelligibility \cite{Taal2011stoi}, Extended STOI \cite{Jensen2016estoi} to measure the intelligibility of the generated samples, Perceptual Evaluation of Speech Quality (PESQ) \cite{Rix2001pesq} (narrow and wide band) to measure the perceptual quality, and Word Error Rate (WER), which we measure with the SOTA ASR network Whisper\cite{radford2023whisper}. 
As expected, \mts is not as accurate as video-to-speech, but the performance is comparable, suggesting that there is sufficient audio signal in the 3D to be useful for our downstream task.
Note that \mts is, however, more accurate than Choi et al.'s original (but not the fine-tuned) model.
We experiment with the following input spaces for \mts: selection of mouth vertices $\verts_{m \in \mouthset}$ (\textit{mouth2s}), all vertices $\verts$ (\textit{face2s}), and FLAME expression parameters $\expparam$ (\textit{exp2s}) and report the results in \tabref{tab:mesh2speech_ablation_public}.
Remarkably, the model which takes global expression vectors, performs slightly better than the selection of mouth vertices $\verts_{m \in \mouthset}$. 
Despite that, we choose \textit{mouth2s} to be our final model, 
as it results in better lip-sync supervision (discussed in Sec.~\ref{sec:exp_speech_driven} and \tabref{tab:other_conditions}).
Please refer to the Sup.~Video for audible comparison and Sec.~\ref{sec:sup_mts} in \supmat for further discussion on performance.
\qheading{\maybedel{Analysis-by-audio-synthesis.}}
\maybedel{We verify if \mts is suitable as an optimizable loss function. Section \ref{sec:sup_mts_audio_synth} in \supmat formulates an optimization that optimizes FLAME expression vectors to produce a target audio. }

\begin{table}[t]
\centering
\resizebox{0.49\textwidth}{!}{
\begin{tabular}{c|l|ccccc}
\toprule
  Modality  &               Model:  & STOI $\uparrow$ & ESTOI $\uparrow$ & PESQ-WB $\uparrow$ & PESQ-NB $\uparrow$ & WER $\downarrow$ \\
\midrule
\multirow{3}{*}{\shortstack{Mesh-to-speech  \\ (\mts)}} &    exp2speech $\expparam$ &           0.502 &   \textit{0.273} &     \textit{1.257} &     \textit{1.468} &   \textit{0.648} \\
 &   face2speech $ \verts $ &           0.458 &            0.203 &              1.225 &              1.437 &            0.953 \\
 &  \textbf{mouth2speech} $ \verts_{m \in \mouthset} $&  \textit{0.506} &            0.272 &              1.246 &              1.457 &            0.678 \\
\midrule
\multirow{2}{*}{\shortstack{Video-to-speech \\ (\svts) }}
& Choi et al.  (finetuned) &  \textbf{0.555} &   \textbf{0.348} &     \textbf{1.281} &     \textbf{1.511} &   \textbf{0.437} \\
& Choi et al.  (orig) &           0.376 &            0.141 &              1.126 &              1.313 &            1.011 \\
\bottomrule
\end{tabular}

}
\caption{{\bf Quantitative comparison} of video-to-speech and 
forms of \mts (\textit{mouth2speech}, \textit{face2speech} and \textit{exp2speech}).
}
\label{tab:mesh2speech_ablation_public}
\end{table}

\subsection{Speech-Driven Facial Animation}
\label{sec:exp_speech_driven}

\qheading{Implementation details.} \model is trained using the Adam optimizer \cite{Kingma2015} ($lr=1e-4$) for 260 epochs with batch size 48, sequence length of 70 frames, and $D=1000$ diffusion steps with a linear noise schedule. The transformer has 8 layers with 4 heads and feature dim $f=128$. 

\qheading{Baselines.}
We compare \model to other speech-driven avatar methods that are also trained on pGT and which output 3DMM parameters (as opposed to vertices).
We reimplemented and trained two SOTA methods, namely FlameFormer* and FlameSelfTalk* (adaptations of FaceFormer \cite{fan2022faceformer} and  \tdv{SelfTalk \cite{peng2023selftalk}}), Media2Face* \cite{zhao2024media2face} and DiffPoseTalk* \cite{sun2024diffposetalk} (see \supmat for details). 
Unless stated otherwise, we keep the pretrained Wav2Vec2 weights frozen. 
Models that finetune it are denoted with suffix -T (T for trainable).
Reimplemented methods are marked with asterisks*.

\qheading{Evaluation protocol.}
For each test audio sequence, we generate S=32 outputs. 
Each of these is initialized with a different random noise in the case of diffusion models, or randomly sampled subject-id condition 
in case of 
FlameFormer.
Then we compute the following evaluation metrics, averaged out across the S outputs (unless stated otherwise).

\qheading{Lip Vertex Error.}
Following \cite{richard2021meshtalk, sun2024diffposetalk, xing2023codetalker} we report LVE, which calculates the maximum L2 error of all lip vertices for each frame and then computes the average over all frames.

\qheading{Dynamic Time Warping.}
We also report the DTW proposed by Thambiraja et al.~\cite{bala2022imitator}. First, distances between the mid-points of lower and upper lips are calculated for both the predictions and GT. The two resulting time series are then used to compute the DTW distance. 

\qheading{Lip Correlation Coefficients.} 
\del{
We argue that the above metrics are not comprehensive enough as LVE only operates over a sparse set of lip vertices and does not model the temporal aspect very well as each frame is handled separately
and DTW only models the lower-upper lip distance. 
A metric that consideres the predictions as time series is needed.
We compute Pearson and Concordance correlation coefficients between the time series of every vertex coordinate in the mouth-region and average the coefficients. 
}
\tdv{LVE is limited to sparse vertices and ignores temporal dynamics, while DTW reduces lip motion to a single distance. To complement these, 
we treat each mouth-region vertex coordinate as a time series and compute Pearson (synchrony) and Concordance (synchrony and bias/scale agreement) correlation coefficients between pGT and predictions.
}

\qheading{Face dynamic deviation.}
\del{An important aspect of facial animation, is the expressiveness of the upper face vertices.}
Following CodeTalker \cite{xing2023codetalker}, we report Face Dynamics Deviation (FDD), which measures the difference between the temporal std.~ deviation of the pGT and the predicted vertices, averaged over a set of vertices. We report upper-FDD (FDD-U) and lip-FDD (FDD-L). 

\qheading{Sample diversity.}
One of the most important aspects of stochastic models is that they generate diverse animations for the same input. 
Remarkably, previous work has not analyzed this aspect \cite{stan2023facediffuser, sun2024diffposetalk, zhao2024media2face}.
We compute standard deviations of vertex distances between the pGT and the predictions for the same input at frame $t$.
Formally, given a tensor of per-vertex L2 distances $\mathbf{D} \in \mathbb{R}^{N \times S\ \times T \times \numverts}$, with $N$ being the size of the test dataset and $S=32$ is the number of sampled outputs per input. 
We compute the standard deviation along the sample dimension $S$. Other dimensions are subsequently averaged. 
We measure sample diversity for %
the upper face region (S-Div-U) and the lip region (S-Div-L).

\qheading{Temporal diversity.}
We also report the temporal diversity, where the standard deviation is computed over the temporal dimension of $\mathbf{D} \in \mathbb{R}^{N \times S\ \times T \times \numverts}$ (and averaged across the rest). Again, we calculate the diversity for both lip (T-Div-L) and upper face regions (T-Div-U).

\subsubsection{Quantitative evaluation}

\begin{table*}[t]
\centering
\resizebox{0.99\textwidth}{!}{
\begin{tabular}{c|c|c|cccc|cc|cc|cc}

\specialrule{.15em}{.1em}{.1em} 
  \multicolumn{3}{c|}{ } & \multicolumn{4}{c}{Lip-Sync} & \multicolumn{2}{|c}{Upper-face Diversity [cm]}  & \multicolumn{2}{|c}{Lip Diversity [cm]} & \multicolumn{2}{|c}{Face Dynamic Dev. [cm]} \\
\toprule
    Experiment    & Input     & Model     & LVE [cm] $\downarrow$       &  L-CCC $\uparrow$  & L-PCC $\uparrow$   & DTW [cm] $\downarrow$    & S-DIV-U  $\uparrow$          & T-DIV-U  $\uparrow$        & S-DIV-L $\downarrow$        & T-DIV-L  $\downarrow$     & T-FDD-U $\downarrow$          & T -FDD-L $\downarrow$     \\
\specialrule{.15em}{.1em}{.1em} 

\multirow{4}{*}{\shortstack{(1) THUNDER and \\ mesh-to-speech \\ input space }} 

& \multirow{4}{*}{\shortstack{ \tdv{audio only}}}  
  &THUNDER w/o m2s              &                   0.879 &              0.359 &              0.568 &               0.329 &             \textbf{0.0419} &            \textbf{0.044} &                        0.21 &                   0.254 &            \textit{0.0118} &                    0.0932  \\
& &THUNDER w/ face2s            &           \textit{0.804}&     \textit{0.411} &     \textit{0.633} &      \textbf{0.285} &                      0.0297 &           \textit{0.0409} &              \textbf{0.128} &          \textit{0.237} &            \textbf{0.0117} &           \textit{0.0827}  \\
& &THUNDER w/ exp2s             &                    0.83 &              0.362 &               0.63 &               0.296 &             \textit{0.0404} &                    0.0398 &                       0.176 &          \textbf{0.228} &                   0.0125   &                    0.0939  \\
& &\textbf{THUNDER w/ mouth2s}  &          \textbf{0.802} &     \textbf{0.426} &     \textbf{0.639} &       \textit{0.29} &                      0.0322 &                      0.04 &              \textit{0.134} &                   0.241 &                     0.0122 &           \textbf{0.0806}  \\
\midrule 

\multirow{2}{*}{\shortstack{(2) THUNDER and \\ trainable audio enc.}} 

& \multirow{2}{*}{\shortstack{ \tdv{audio only}}}  
  &THUNDER-T w/o m2s            &                   0.723 &              0.428 &              0.623 &               0.266 &                       0.020 &           \textbf{0.0411} &             \textbf{0.0656} &                   0.216 &                      0.0118 &                   0.0811  \\
& &\textbf{THUNDER-T w/ mouth2s}    &      \textbf{0.709} &     \textbf{0.445} &      \textbf{0.66} &      \textbf{0.256} &              \textbf{0.021} &                     0.039 &                      0.0669 &          \textbf{0.202} &                      0.0118 &          \textbf{0.0788}  \\

\specialrule{.15em}{.1em}{.1em} 

\multirow{4}{*}{\shortstack{(3) Mesh-to-speech \\ with other diffusion-based \\ methods which use additional \\ input conditions}} 
& \multirow{2}{*}{\shortstack{\tdv{audio}, \\ \tdv{image feature}}} 
  &Media2Face*                  &         \textit{0.96}  &     \textit{0.308} &     \textit{0.531} &      \textit{0.363} &             \textbf{0.0163} &           \textit{0.0459} &              \textit{0.108} &          \textit{0.256} &             \textbf{0.0105} &           \textit{0.0899}  \\
& &Media2Face* w/ m2s           &        \textbf{0.815}  &     \textbf{0.428} &     \textbf{0.609} &      \textbf{0.282} &             \textit{0.0128} &           \textbf{0.0586} &             \textbf{0.0371} &          \textbf{0.233} &             \textit{0.0154} &            \textbf{0.083}  \\

\cmidrule{2-13} 

& \multirow{2}{*}{\shortstack{\tdv{audio}, \\ \tdv{speaker style feat.}}}  
  &DiffPoseTalk*                &          \textit{0.68} &     \textit{0.464} &     \textit{0.612} &      \textbf{0.267} &              \textbf{0.025} &           \textbf{0.0379} &              \textit{0.117} &          \textit{0.223} &             \textbf{0.0101} &           \textbf{0.0776}  \\
& &DiffPoseTalk* w/ m2s         &         \textbf{0.651} &     \textbf{0.469} &     \textbf{0.653} &      \textit{0.268} &             \textit{0.0172} &           \textit{0.0377} &             \textbf{0.0604} &          \textbf{0.204} &             \textit{0.0117} &           \textit{0.0813}  \\

\specialrule{.15em}{.1em}{.1em} 
\multirow{4}{*}{\shortstack{ (4) Mesh-to-speech \\ with deterministic methods \\ and discrete \\ input conditions}} 
& \multirow{2}{*}{\shortstack{\tdv{audio}, \\ \tdv{one-hot speaker ID}}} 
  &FlameFormer*                 &       \textit{0.809}   &     \textit{0.368} &      \textit{0.57} &      \textit{0.291} &             \textbf{0.0271} &           \textit{0.0372} &              \textit{0.132} &          \textit{0.239} &             \textit{0.0117} &           \textit{0.0909}  \\
& &FlameFormer* w/ m2s          &         \textbf{0.794} &     \textbf{0.411} &     \textbf{0.614} &      \textbf{0.265} &             \textit{0.0263} &           \textbf{0.0393} &              \textbf{0.111} &          \textbf{0.219} &             \textbf{0.0107} &           \textbf{0.0715}  \\

\cmidrule{2-13} 
& \multirow{2}{*}{\shortstack{\tdv{audio}, \\ \tdv{one-hot speaker ID}}} 
   & \tdv{FlameSelfTalk*}                   &    \textit{0.695} &     \textit{0.504} &     \textit{0.698} &      \textbf{0.243} &              \textit{0.0275} &           \textit{0.0372} &              \textit{0.108} &          \textit{0.201} &            \textit{0.00983} &           \textit{0.0692} \\
& & \tdv{FlameSelfTalk* w/ m2s}        &    \textbf{0.674} &     \textbf{0.518} &     \textbf{0.705} &      \textit{0.243} &               \textbf{0.029} &            \textbf{0.039} &              \textbf{0.108} &          \textbf{0.193} &            \textbf{0.00965} &           \textbf{0.0674} \\

\specialrule{.15em}{.1em}{.1em} 
\end{tabular}

}
\vspace{-0.2cm}
\caption{
{\bf Quantitative evaluation} 
\tdv{on THUNDERSET}
\del{of \model, other models and ablations. }
{\bf (1) Mesh-to-speech input space.} 
All of the \mts variants are applicable. \model with \textit{mouth2s} results in the best lip-sync performance (LVE, L-CCC, L-PCC), which is why we select it as the final \model model. \textit{exp2s} is applicable, too as it scores best on DTW and has good diversity (S-DIV-U), while \textit{face2s} scores the worst on diversity (S-DIV-U, T-DIV-U).
{\bf (2) Trainable audio encoder.} \model-T results in improved lip-sync metrics (LVE, L-CCC, L-PCC and DTW), but at slight expense of diversity of outputs (mainly upper face diversity S-DIV-U and T-DIV-U) compared to \model which does not train the audio encoder. 
\tdv{ {\bf (3) Mesh-To-Speech with diffusion-based methods.}
Training Media2Face* (with image conditioning) and DiffPoseTalk* (style conditioning) 
follows the same trend - increase in lip-sync quality and reduction in upper-face diversity.
}
\tdv{ {\bf (4) Mesh-To-Speech with deterministic methods.} 
We train FlameFormer* and SelfTalk* with \mts and find that it improves the lip-sync related metrics.
}
\del{{\bf (3) Mesh-To-Speech applied to other methods.} The application of \mts to our reimplementations of Media2Face* (diffusion with image conditioning), DiffPoseTalk* (diffusion with style feature conditioning) and FlameFormer (deterministic with subject ID conditioning) improves lip-sync metrics at slight expense of expression diversity. 
Please note that DiffPoseTalk* and Media2Face* (both with and without \mts) have a considerably lower diversity than \model. This happens due to the additional conditioning inputs which narrow down the output distribution space. This also manifests itself with better lip-sync performance.
}
}
\label{tab:other_conditions}
\end{table*}
Tab.~\ref{tab:other_conditions} reports the test-set metrics computed for models trained on THUNDERSET 
\tdv{and Tab.~\ref{tab:TFHP_main_paper} does so for models trained on TFHP}. 
\tdv{We analyze the following:}

\tdv{\qheading{Does \mts improve THUNDER?} 
Yes, models with \mts improve all lip-sync metrics at slight expense of diversity (see \tabref{tab:other_conditions}, experiments~1 and 2)}.

\tdv{\qheading{Does the input space for \mts matter?} While all models improve lip-sync metrics, \emph{mouth2speech} has the strongest effect on lip-sync metrics, likely thanks to its localized effect on the mouth region  (\tabref{tab:other_conditions}, experiment 1).}

\tdv{\qheading{Should the audio encoder be finetuned?} Trainable encoder (what most methods use) results in much lower diversity, presumably due to overfitting to the training voices. 
Frozen encoder results in inferior lip-sync but higher diversity. 
\mts improves lip-sync in both cases. 
\mts applied with a frozen encoder leads to a dramatically improved lip-sync while retaining some of the upper-face diversity (\tabref{tab:other_conditions}, exp.~1,2).
}

\tdv{\qheading{Does \mts improve other methods?} Yes, we apply \mts to train the deterministic FlameFormer, SelfTalk and stochastic diffusion-based Media2Face and DiffPoseTalk and show improvement on the lip-sync metrics
(\tabref{tab:other_conditions}, exp.~3,4)
.
}

\tdv{\qheading{Does \model generalize to other datasets?} Yes. We train and test \mts and \model on the challenging TFHP (the original DiffPoseTalk dataset). \model outperforms original DiffPoseTalk (see \tabref{tab:TFHP_main_paper} and Sec.~\ref{sec:otherdatasets} in \supmat)}

\tdv{\qheading{Can \model generate head pose motion?} Yes. We train a \model model with head pose on TFHP and show improvement over DiffPoseTalk (see \tabref{tab:TFHP_main_paper} and Sec.~\ref{sec:otherdatasets}). 
}

\tdv{\qheading{Does \mts help with editing methods?} Yes. We feed images of different emotions to our M2F. M2F w/ \mts improves lip-sync (see Sec.~\ref{sec:disentanglement} in \supmat)}

\begin{table*}[t]
\centering
\resizebox{\textwidth}{!}{
\begin{tabular}{c|c|llll|ll|ll|ll|ll}

\specialrule{.15em}{.1em}{.1em} 
  \multicolumn{2}{c|}{ } & \multicolumn{4}{c}{Lip-Sync} & \multicolumn{2}{|c}{Upper-face diversity [mm] }  & \multicolumn{2}{|c}{Lip diversity  [mm]} & \multicolumn{2}{|c}{Face Dynamic Dev. [mm]} & \multicolumn{2}{|c}{Head Pose}\\
\toprule
     Name    & Backbone         &    LVE [cm] $\downarrow$ & L-CCC $\uparrow$ & L-PCC $\uparrow$ & DTW [cm] $\downarrow$ & S-DIV-U $\uparrow$ & T-DIV-U   $\uparrow$ & S-DIV-L  $\downarrow$ & T-DIV-L  $\downarrow$ & FDD-U $\downarrow$ & FDD-L $\downarrow$  & BA $\times10^{-3}$ $\uparrow$ & DIV $\uparrow$ \\
\specialrule{.15em}{.1em}{.1em}

   THUNDER-F w/o m2s        &   Wav2Vec2 frozen     &     \textit{1.2} &   \textit{0.285} &   \textit{0.388} &    \textit{2.26} &         \textbf{0.258} &         \textbf{0.349} &           \textit{2.6} &        \textit{3.58} &          \textbf{0.0649} &          \textbf{0.76} &           \textit{6.02} &     \textit{3.07} \\
   THUNDER-F                &   Wav2Vec2 frozen      &    \textbf{1.07} &   \textbf{0.298} &   \textbf{0.435} &    \textbf{1.95} &         \textit{0.203} &         \textit{0.321} &           \textbf{1.5} &        \textbf{2.98} &          \textit{0.0707} &          \textit{0.95} &          \textbf{6.13} &     \textbf{3.78} \\

\midrule
              DiffPoseTalk   &   HuBERT trainable    &     \textit{1.01} &   \textit{0.414} &   \textit{0.541} &    \textit{1.85} &         \textbf{0.247} &         \textbf{0.352} &          \textit{1.58} &        \textit{3.02} &                   0.0641 &         \textit{0.618} &                         5.54 &              2.16 \\
          THUNDER-T w/o m2s  &   Wav2Vec2 trainable     &               1.12 &            0.358 &            0.488 &    \textit{1.86} &                  0.241 &         \textit{0.351} &                   1.87 &               3.21 &          \textit{0.0597} &                  0.661 &       \textbf{6.25} &     \textit{3.49} \\
                  THUNDER-T  &   Wav2Vec2 trainable     &   \textbf{0.987} &   \textbf{0.422} &    \textbf{0.55} &    \textbf{1.81} &         \textit{0.244} &                  0.341 &          \textbf{1.35} &          \textbf{2.89} &          \textbf{0.0578} &          \textbf{0.55} &          \textit{5.77} &     \textbf{4.83} \\
\specialrule{.15em}{.1em}{.1em} 
\end{tabular}
} 
\vspace{-0.3cm}
\caption{ 
{ \bf Quantitative evaluation on TFHP. }
We compare \model, and \model w/o \mts and the original DiffPoseTalk release (all trained on TFHP). Consistently to our other experiments, \model results in superior lip-sync performance over \model w/o \mts and DiffPoseTalk, at a slight trade-off of diversity metrics. Notably, when trained with head pose, all \model variants outperforms DiffPoseTalk on head pose metrics - beat alignment (BA) and head pose diversity (DIV). 
Refer to Sup.~Video and PDF for qualitative results.
} 
\label{tab:TFHP_main_paper}
\vspace{-0.4cm}
\end{table*}

\del{
(1) The input space for the \mts loss, 
(2) the trainability of the audio encoder (Tab.~\ref{tab:other_conditions})
, and (3) the weight of the \mts loss (see Sec.~\ref{sec:extened_ablation} in \supmat and video).
}
\del{\qheading{\mts loss with other methods.}}
\del{Tab.~\ref{tab:other_conditions} shows that \mts is applicable to other methods.
We re-implement three  methods that incorporate additional conditioning mechanisms and train them with and without our \mts loss.
Our new loss improves performance on the lip-sync metrics. 
For visual results, refer to Sup.~Video.
}

\maybedel{
\qheading{Comparison to other methods.} Tab.~\ref{tab:talkinghead} in \supmat includes a comparison of \model to other recent methods. 
}

\del{
\qheading{Disentanglement effect.} 
To test whether \mts helps preserve the effect of good lip-sync in the presence of other \textit{editing conditions}, we additionally run our reimplementations of Media2Face with conditioning inputs extracted from emotional images. 
See Sec.~\ref{sec:disentanglement} in \supmat
}
\del{
\qheading{Other datasets.} 
To verify, whether \model is dataset-agnostic, we also trained and tested the method on TFHP. We report the results in Sec.~\ref{sec:otherdatasets} in \supmat.
}

\begin{figure}[b]
    \vspace{-0.75cm}
    \offinterlineskip
    \centering
    \includegraphics[trim=0.5cm 0.7cm 0.7cm 1.5cm, clip, width=1.05\columnwidth]{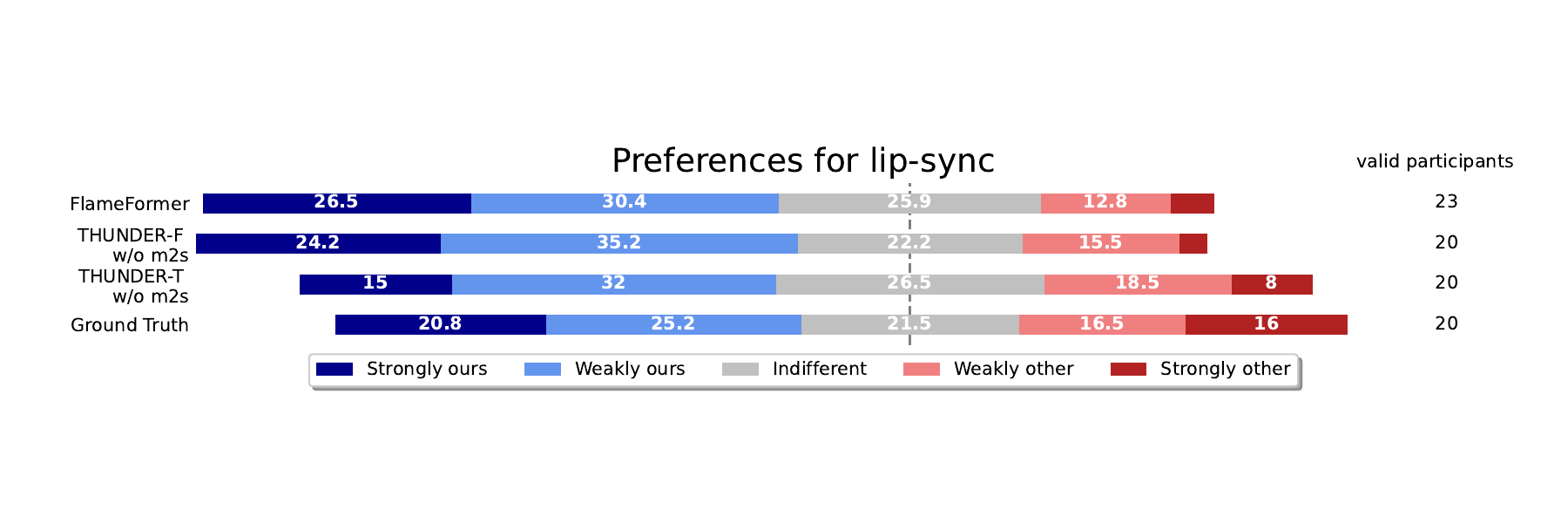}
    \vspace{-0.5cm}
    \caption{
       \textbf{Perceptual study of \model.} 
       We compare \model-F (-F for frozen backbone) with methods having both a trainable audio encoder (THUNDER-T w/o m2s) and frozen encoders (FlameFormer-F, THUNDER-F w/o m2s), and GT. 
       The participants prefer \model-F's lip-sync over that of the other models. 
       \del{, but does not quite reach the fidelity of GT.}
       Remarkably, the participants also have  slight preference for \model-F over GT, which suggests that the application of \mts helps \model saturate the quality of GT.
    }
    \label{fig:main_perceptual_thunderF}
\end{figure}

\qheading{Perceptual study.}
We conduct a perceptual study in which we evaluate \model and the effect of mesh-to-speech. 
Specifically, we compare to the deterministic FlameFormer, \model with and without \mts, and p-GT. 
The participants are shown two videos side-by-side and asked to rate \tdv{three aspects of the (lip-sync, dynamism, and realism)} animations on a five-point Likert scale (strong/weak preference for A or B or indifferent). 
\del{We ask the participants to rate three aspects of the animation: lip-sync quality, dynamism, and realism. }
We find that participants prefer \model over other methods. \figref{fig:main_perceptual_thunderF} reports the results for \model for lip-sync. 
For the remaining results of the study and further discussion, see Sec.~\ref{sec:supmat_perceptual} in \supmat

\subsubsection{Qualitative evaluation}

\begin{figure}[t]
    \offinterlineskip
    \centering
    \includegraphics[width=1.\columnwidth]{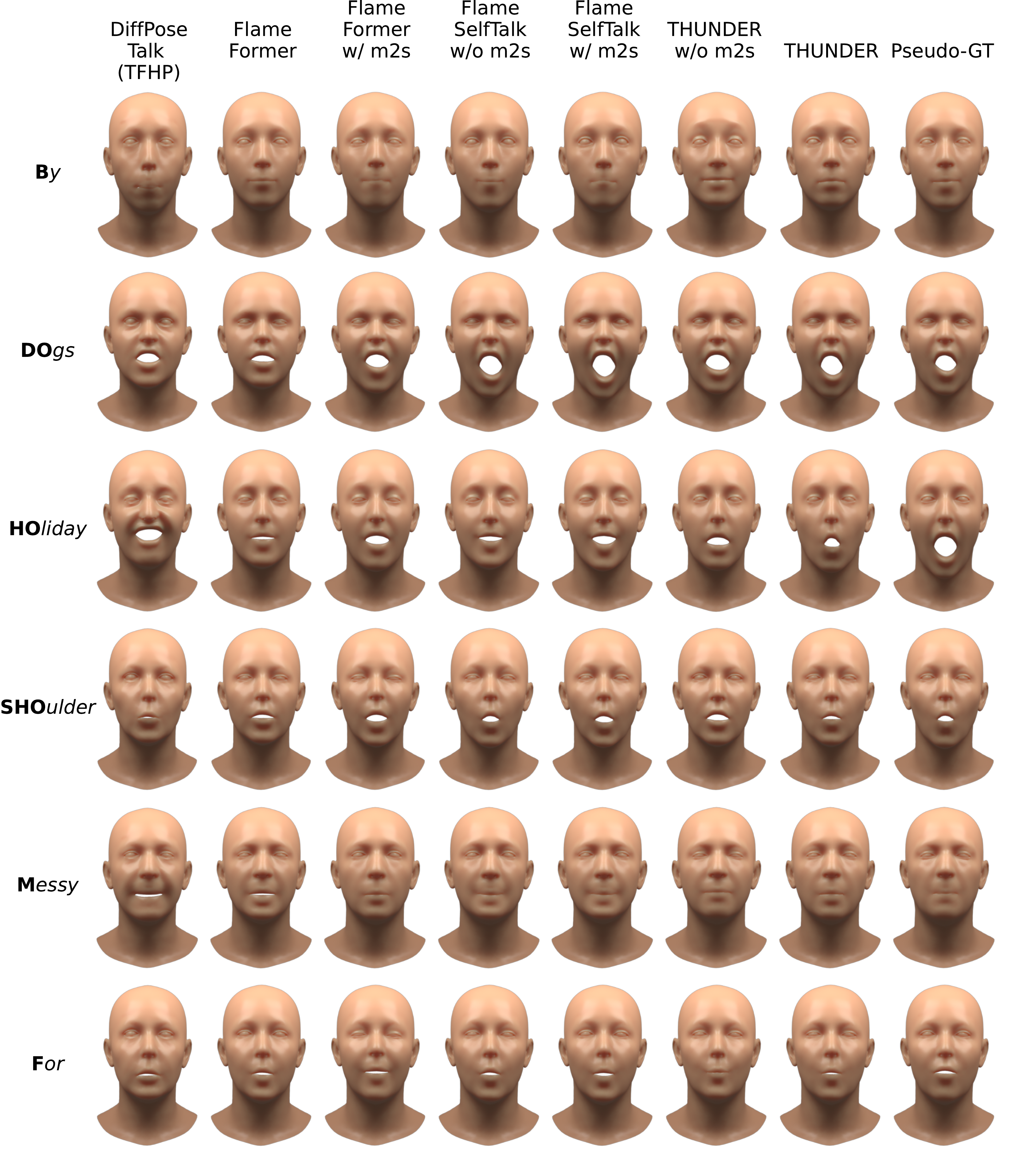}
    \vspace{-0.3cm}
    \caption{
        {\bf Qualitative comparison on THUNDERSET.} This figure shows the comparison between baselines, our model and GT for selected utterances. 
        Note that DiffPoseTalk was trained on TFHP. Supplemental PDF and video contain more qualitative comparisons.
    }
    \vspace{-0.7cm}
    \label{fig:comparison}
\end{figure}

We refer the reader to the supplemental video for a detail comparisons of the animations, which
demonstrates the superiority of our method qualitatively. 
\figref{fig:comparison} shows the comparison of our method with the baselines for selected utterances. 
Additional qualitative evaluation can be found in \supmat.

\subsection{Limitations}
\model has the following limitations. 
The absence of teeth and tongue inherently creates ambiguity, making it more difficult to produce high quality audio compared to silent video-to-speech. 
Further, 
pseudo-GT is not a perfect reconstruction but has occasional artifacts or inaccuracies.  
Also, our current \mts architecture is fully deterministic and hence is not completely capable of capturing the prediction ambiguities. 
All of these may result in suboptimal effects of the audio cycle-consistency loss. 
Regardless of these shortcomings, our experiments demonstrate the benefits of our method.
Employing a large-scale dataset of high quality 3D scans, a face model that models teeth and tongue, and upgrade \mts to a stochastic model are likely to boost the benefits of the \mts loss. 
Finally, the inference process of the diffusion model is currently computationally intensive. 
Future work should address this by employing more efficient solvers such as DPM++ \cite{lu2022dpmpp}.

\section{Conclusion} 

In this paper, we introduced \model, a 3D speech-driven avatar generation system with a novel paradigm of supervision via 
analysis-by-audio-synthesis.
\model has two key components.
First, drawing inspiration from silent video-to-speech methods, we define a new mesh-to-speech task and train a model that regresses speech from facial animation. 
Second, we incorporated mesh-to-speech into a diffusion-based 3D speech-driven avatar system, creating the first-of-its-kind audio-based self-supervision loop that helps ensure that the produced lip animation is plausible for the spoken audio. 
Our extensive quantitative, perceptual, and qualitative experiments demonstrate that \model achieves a significant improvement in lip animation quality for diffusion-based 3D speech-driven avatars while still producing a diverse set of facial animations. 
Furthermore, we have demonstrated that the application of our mesh-to-speech loss improves the lip-sync quality of other talking head avatar systems.
Finally, we believe that our novel audio-based self-supervision paradigm can impact other applications of 3D head avatars, such as video-based reconstruction, automatic postprocessing of 3D facial animation, or as a quality measure of talking head avatars. 

{

\footnotesize
\textcolor{white}{ }

\qheading{Acknowledgments.} We thank Raja Bala, Hiro Takeda, Brandon Smith, Ganesh Iyer, Chinghang Chen, Alex Vorobiov and Betty Mohler for helpful discussions. 
This project was supported by an Amazon Research Award. %

\qheading{Disclosure:} While MJB is a co-founder and Chief Scientist at Meshcapade, his contribution was performed at, and funded by, the MPG.
}
{
    \small
    \bibliographystyle{ieeenat_fullname}
    \bibliography{main}
}
\clearpage
\setcounter{page}{1}

\ifarxiv 
\appendix 
\else 
\maketitlesupplementary
\fi

\section{Additional Experiments}

\subsection{Mesh-to-Speech }
\label{sec:sup_mts}

\subsubsection{Per-Dataset Performance Analysis.}
\label{sec:sup_mts_per_dataset}
The performance of both mesh-to-speech and silent-video-to-speech models varies dramatically depending on the dataset. Datasets with limited number of utterances and vocabulary such as RAVDESS or GRID are less challenging, and the models can regress relatively accurately, sometimes to the point of perfect intelligibility. On the more challenging TCD-TIMIT dataset, which contains more complex vocabulary, words are often missed. 
However, despite the relatively high word error rate (WER) on TCD-TIMIT, the produced sounds remain plausible given the input animation. 
Please see \tabref{tab:mesh2speech} for per-dataset breakdown and refer to the supplementary video for an audible comparison.

\begin{table}[h] %
\centering
\resizebox{0.49\textwidth}{!}{
\begin{tabular}{l|l|ccccc}
\specialrule{.15em}{.1em}{.1em}
DATASET                   &       Model:  &  STOI $\uparrow$ &  ESTOI $\uparrow$ &  PESQ-WB $\uparrow$ &  PESQ-NB $\uparrow$ &   WER $\downarrow$ \\
\specialrule{.15em}{.1em}{.1em} 
\multirow{5}{*}{GRID}      
&  exp2speech $\expparam$ &           0.533 &            0.290 &     \textit{1.298} &     \textit{1.552} &   \textit{0.553} \\
&  face2speech $ \verts $ &           0.482 &            0.211 &              1.261 &              1.513 &            0.783 \\
&  \textbf{mouth2speech} $ \verts_{m \in \mouthset} $  &  \textit{0.538} &   \textit{0.295} &              1.288 &              1.539 &            0.559 \\
& Choi et al. (finetuned) &  \textbf{0.590} &   \textbf{0.371} &     \textbf{1.324} &     \textbf{1.597} &   \textbf{0.428} \\
&             Choi et al. &           0.426 &            0.158 &              1.137 &              1.345 &            1.202 \\
\midrule 

\multirow{5}{*}{RAVDESS}   
&  exp2speech $\expparam$ &           0.460 &            0.284 &              1.159 &              1.262 &            0.145 \\
&  face2speech $ \verts $  &           0.474 &            0.293 &     \textit{1.164} &     \textit{1.266} &            0.173 \\
&   \textbf{mouth2speech} $ \verts_{m \in \mouthset} $ &  \textit{0.510} &   \textit{0.328} &              1.161 &              1.262 &   \textit{0.131} \\
& Choi et al. (finetuned) &  \textbf{0.548} &   \textbf{0.375} &     \textbf{1.174} &     \textbf{1.277} &   \textbf{0.060} \\
&             Choi et al. &           0.452 &            0.245 &              1.074 &              1.149 &            0.774 \\
\midrule %

\multirow{5}{*}{TCD-TIMIT} 
&  exp2speech $\expparam$ &  \textit{0.409} &   \textit{0.206} &     \textit{1.146} &              1.249 &            0.929 \\
&  face2speech $ \verts $  &           0.368 &            0.139 &              1.117 &              1.228 &            1.431 \\
&   \textbf{mouth2speech} $ \verts_{m \in \mouthset} $&           0.389 &            0.174 &              1.131 &              1.243 &            1.070 \\
&Choi et al. (finetuned) &  \textbf{0.436} &   \textbf{0.257} &     \textbf{1.168} &     \textbf{1.293} &   \textbf{0.579} \\
&            Choi et al. &           0.172 &            0.041 &              1.103 &     \textit{1.254} &   \textit{0.641} \\
\specialrule{.15em}{.05em}{.05em} 

\multirow{5}{*}{COMBINED}       
&  exp2speech $\expparam$ &           0.502 &   \textit{0.273} &     \textit{1.257} &     \textit{1.468} &   \textit{0.648} \\
&   face2speech $ \verts $  &           0.458 &            0.203 &              1.225 &              1.437 &            0.953 \\
&     \textbf{mouth2speech} $ \verts_{m \in \mouthset} $&  \textit{0.506} &            0.272 &              1.246 &              1.457 &            0.678 \\
& Choi et al. (finetuned) &  \textbf{0.555} &   \textbf{0.348} &     \textbf{1.281} &     \textbf{1.511} &   \textbf{0.437} \\
& Choi et al. (orig) &           0.376 &            0.141 &              1.126 &              1.313 &            1.011 \\

\specialrule{.15em}{.1em}{.1em} 
\end{tabular}

}
\caption{{\bf Quantitative comparison of mesh-to-speech with silent-video-to-speech.} 
While Choi et al.~outperforms \model in most cases, the performance is comparable.
The performance of both methods depends dramatically on the test dataset. 
Inputs from datasets with limited vocabulary (GRID, RAVDESS) tend to produce lower word error rates than datasets with richer vocabulary (TCD-TIMIT). 
}
\label{tab:mesh2speech}
\end{table}

\subsection{Speech-Driven Facial Animation}

\subsubsection{Experiments on TFHP}
\label{sec:otherdatasets}

\begin{table*}[t]
\centering
\resizebox{\textwidth}{!}{
\begin{tabular}{c|c|c|c|llll|ll|ll|ll|ll|ll}

\specialrule{.15em}{.1em}{.1em} 
  \multicolumn{4}{c|}{ } & \multicolumn{4}{c}{Lip-Sync} & \multicolumn{2}{|c}{Upper diversity [mm]}  & \multicolumn{2}{|c}{Lip diversity [mm]} & \multicolumn{2}{|c}{Face Dynamic Dev. [mm]} &  \multicolumn{2}{|c}{Head Pose} & \multicolumn{2}{|c}{Chae et.~al \cite{chae2025perceptually}}\\
\toprule
    Experiment  &   Name    & Backbone   & Input                    &    LVE [mm] $\downarrow$ & L-CCC $\uparrow$ & L-PCC $\uparrow$ & DTW [mm] $\downarrow$ & S-DIV-U $\uparrow$ & T-DIV-U $\uparrow$ & S-DIV-L $\downarrow$ & T-DIV-L $\downarrow$ & T-FDD-U $\downarrow$ & T-FDD-L $\downarrow$ & BA $\times 10^{-3}$  $\uparrow$ & DIV $\uparrow$ & MTM [ms] $\downarrow$ & PLRS $\uparrow$ \\
\specialrule{.15em}{.1em}{.1em}

\multirow{5}{*}{\shortstack{(1) models \\ trained  without \\ head pose}} 
 & THUNDER-F w/o m2s        &   Wav2Vec2 frozen    &    shape, audio    &    \textit{1.21} &   \textit{0.316} &    \textit{0.42} &    \textit{2.37} &         \textbf{0.299} &         \textbf{0.386} &           \textit{2.7} &        \textit{3.71} &          \textit{0.0739} &          \textit{0.79}        & n/a  & n/a   & \textit{125.141} &  \textit{0.175}\\
 & THUNDER-F                &   Wav2Vec2 frozen    &    shape, audio    &   \textbf{1.05}  &   \textbf{0.317} &   \textbf{0.441} &    \textbf{1.93} &         \textit{0.195} &         \textit{0.328} &           \textbf{1.5} &        \textbf{3.12} &          \textbf{0.0627} &         \textbf{0.759}        & n/a  & n/a   & \textbf{103.185} &  \textbf{0.269} \\

\cmidrule{2-18}
 & DiffPoseTalk             &   HuBERT trainable    & shape, audio       &     \textit{1.02} &    \textit{0.41} &   \textit{0.526} &             1.84 &         \textbf{0.272} &         \textbf{0.383} &                    1.7 &                  2.98 &                   0.0764 &                 0.679      & n/a  & n/a  &  \textit{93.485} &  \textit{0.265} \\
 &     THUNDER-T  w/o m2s   &   Wav2Vec2 trainable  &  shape, audio      &              1.03 &            0.388 &            0.525 &    \textit{1.84} &         \textit{0.265} &         \textit{0.364} &          \textit{1.58} &        \textit{2.95} &          \textit{0.0679} &         \textit{0.609}      & n/a  & n/a  &           94.132 &           0.247 \\
 &    THUNDER-T             &   Wav2Vec2 trainable  & shape, audio       &    \textbf{0.963} &   \textbf{0.415} &    \textbf{0.55} &    \textbf{1.77} &                  0.228 &                  0.341 &          \textbf{1.09} &        \textbf{2.79} &          \textbf{0.0592} &          \textbf{0.59}      & n/a  & n/a  &  \textbf{88.908} &  \textbf{0.329} \\

\specialrule{.15em}{.1em}{.1em} 

\multirow{5}{*}{\shortstack{(2) models \\ trained with \\ head pose}} 
 &  THUNDER-F w/o m2s        &   Wav2Vec2 frozen  &      shape, audio    &     \textit{1.2} &   \textit{0.285} &   \textit{0.388} &    \textit{2.26} &         \textbf{0.258} &         \textbf{0.349} &           \textit{2.6} &        \textit{3.58} &          \textbf{0.0649} &          \textbf{0.76}  &    \textit{6.02} &     \textit{3.07} & \textit{134.403} &  \textit{0.153} \\
 &  THUNDER-F                &   Wav2Vec2 frozen  &      shape, audio    &    \textbf{1.07} &   \textbf{0.298} &   \textbf{0.435} &    \textbf{1.95} &         \textit{0.203} &         \textit{0.321} &           \textbf{1.5} &        \textbf{2.98} &          \textit{0.0707} &          \textit{0.95}  &    \textbf{6.13} &     \textbf{3.78} & \textbf{105.536} &  \textbf{0.292} \\

\cmidrule{2-18} 
 &             DiffPoseTalk   &   HuBERT trainable  &  shape, audio     &     \textit{1.01} &   \textit{0.414} &   \textit{0.541} &    \textit{1.85} &         \textbf{0.247} &         \textbf{0.352} &          \textit{1.58} &        \textit{3.02} &                   0.0641 &         \textit{0.618} &                5.54 &              2.16   &  \textit{90.005} &  \textit{0.251} \\
 &         THUNDER-T w/o m2s  &   Wav2Vec2 trainable&  shape, audio     &               1.12 &            0.358 &            0.488 &    \textit{1.86} &                  0.241 &         \textit{0.351} &                   1.87 &               3.21  &          \textit{0.0597} &                  0.661 &       \textbf{6.25} &     \textit{3.49}  &        101.232 &           0.230 \\
 &                 THUNDER-T  &   Wav2Vec2 trainable&  shape, audio     &   \textbf{0.987} &   \textbf{0.422} &    \textbf{0.55} &    \textbf{1.81} &         \textit{0.244} &                  0.341 &          \textbf{1.35} &           \textbf{2.89} &          \textbf{0.0578} &          \textbf{0.55} &       \textit{5.77} &     \textbf{4.83}  &  \textbf{89.550} &  \textbf{0.320} \\
\specialrule{.15em}{.1em}{.1em}

\multirow{5}{*}{\shortstack{(3) models \\  trained with \\ head pose \\ and style \\ condition}} 
 &   DiffPoseTalk              &   HuBERT trainable   &  audio, shape, style &   \textit{0.897} &            0.439 &   \textit{0.555} &             1.77 &         \textit{0.191} &         \textbf{0.319} &          \textit{1.27} &        \textit{2.86} &                   0.0632 &         \textbf{0.572} &                5.57 &              1.43  &  \textbf{88.647} &           0.256 \\
 &   THUNDER-T w/o m2s         &   Wav2Vec2 trainable&  audio, shape, style  &            0.899 &   \textbf{0.443} &            0.545 &    \textit{1.73} &         \textbf{0.206} &                   0.31 &                   1.45 &                 2.91 &          \textbf{0.0549} &         \textit{0.573} &       \textit{5.72} &      \textbf{3.3} &           94.046 &  \textit{0.261}\\
 &   THUNDER-T                 &   Wav2Vec2 trainable&  audio, shape, style  &   \textbf{0.891} &   \textit{0.442} &   \textbf{0.567} &    \textbf{1.72} &                  0.186 &          \textit{0.31} &         \textbf{0.941} &         \textbf{2.81} &          \textit{0.0598} &                 0.625 &       \textbf{5.98} &     \textit{1.96} &  \textit{89.108} &  \textbf{0.327} \\
\cmidrule{2-18} 
 &  THUNDER-F w/o m2s          &   Wav2Vec2 frozen &  audio, shape, style    &    \textit{1.02} &   \textit{0.309} &   \textit{0.386} &    \textit{2.04} &         \textbf{0.227} &         \textbf{0.323} &          \textit{2.39} &        \textit{3.42} &          \textbf{0.0646} &         \textbf{0.726} &       \textit{5.85} &     \textbf{2.64}  &  \textit{134.639} &  \textit{0.124} \\
 & THUNDER-F                   &   Wav2Vec2 frozen &  audio, shape, style    &   \textbf{0.917} &   \textbf{0.361} &   \textbf{0.465} &    \textbf{1.85} &         \textit{0.181} &         \textit{0.312} &          \textbf{1.49} &        \textbf{2.98} &          \textit{0.0679} &         \textit{0.755} &       \textbf{5.92} &      \textit{2.4}  &  \textbf{103.993} &  \textbf{0.280}\\
\specialrule{.15em}{.1em}{.1em}

\multirow{8}{*}{\shortstack{(4) models \\ trained  with \\ perc. loss \\  Chae et al.~\cite{chae2025perceptually}}} 

 &   THUNDER-T w/o m2s, perc.         &   Wav2Vec2 trainable&  audio, shape  &            1.117 &            0.358 &            0.488 &   \textit{1.864} &                  0.241 &         \textbf{0.351} &                   1.87 &                  3.21 &           \textit{0.060} &         \textit{0.661} &      \textbf{6.25} &             3.49 &          101.232 &           0.230   \\
 &   THUNDER-T w/o m2s, w/ perc.      &   Wav2Vec2 trainable&  audio, shape  &            1.117 &            0.325 &            0.485 &            1.991 &         \textbf{0.246} &         \textit{0.344} &                   1.60 &                  3.00 &                    0.064 &                  0.783 &               5.71 &    \textit{3.99} &          100.658 &  \textbf{0.472} \\
  &   THUNDER-T w/ m2s, w/o perc.     &   Wav2Vec2 trainable&  audio, shape  &   \textbf{0.987} &   \textbf{0.422} &   \textbf{0.550} &   \textbf{1.811} &         \textit{0.244} &                  0.341 &          \textit{1.35} &         \textit{2.89} &           \textbf{0.058} &         \textbf{0.550} &               5.77 &    \textbf{4.83} &  \textit{89.550} &           0.320 \\
  &   THUNDER-T  w/ m2s, perc.         &   Wav2Vec2 trainable&  audio, shape  &   \textit{1.079} &   \textit{0.365} &   \textit{0.537} &            1.869 &                  0.224 &                  0.329 &          \textbf{1.11} &         \textbf{2.74} &                    0.064 &                 0.777 &      \textit{5.92} &             2.88 &  \textbf{88.862} &  \textit{0.469}  \\

\cmidrule{2-18} 

 &   THUNDER-F w/o m2s, perc.         &   Wav2Vec2 frozen&  audio, shape     &              1.2 &            0.285 &            0.388 &             2.26 &                  0.258 &                  0.349 &                    2.6 &                   3.58 &          \textbf{0.0649} &                   0.76 &      \textit{6.02} &              3.07 &          134.403 &           0.153  \\
 &   THUNDER-F w/o m2s, w/ perc.      &   Wav2Vec2 frozen&  audio, shape     &              1.2 &   \textit{0.318} &   \textit{0.462} &             2.27 &          \textit{0.28} &         \textit{0.364} &          \textit{2.07} &          \textit{3.11} &          \textit{0.0652} &         \textit{0.704} &               5.98 &              3.66 &          107.130 &  \textit{0.455}  \\
  &   THUNDER-F w/ m2s, w/o perc.     &   Wav2Vec2 frozen&  audio, shape     &    \textbf{1.07} &            0.298 &            0.435 &    \textbf{1.95} &                  0.203 &                  0.321 &           \textbf{1.5} &          \textbf{2.98} &                   0.0707 &                   0.95 &      \textbf{6.13} &     \textit{3.78} & \textit{105.536} &           0.292  \\
 &   THUNDER-F   w/ m2s, perc.        &   Wav2Vec2 frozen&  audio, shape     &    \textit{1.16} &   \textbf{0.348} &   \textbf{0.498} &    \textit{2.07} &          \textbf{0.33} &         \textbf{0.375} &                   2.13 &                   3.23 &                   0.0751 &         \textbf{0.687} &                5.9 &     \textbf{4.48} &  \textbf{99.495} &  \textbf{0.479} \\

\specialrule{.15em}{.1em}{.1em} 
\end{tabular}
} 
\caption{ 
{ \bf Experiment on TFHP. }
We compare \model-T, \model w/o mesh-to-speech and DiffPoseTalk (all trained on TFHP). Similarly to our other experiments, \model-T results in superior lip-sync metrics, while trading off upper-face diversity.
} 
\label{tab:TFHP}
\end{table*}

To verify whether the performance of \model and mesh-to-speech translates to other types of data, we also conduct a comprehensive evaluation on TFHP (the DiffPoseTalk dataset \cite{sun2024diffposetalk}).
TFHP was reconstructed from YouTube videos.
The dataset contains unscripted audios with unconstrained longer sequences. 
The reconstructions were obtained using MICA \cite{MICA:ECCV2022} and SPECTRE \cite{filntisis2022visual} and look qualitatively very different compared to our EMICA reconstructions of GRID, RAVDESS and TCD-TIMIT datasets. 
Can \model work also in this setting?

We train both \mts and \model on TFHP and compare it to the official DiffPoseTalk release.
To achieve maximum fairness of comparison, we adopt DiffPoseTalk's number of diffusion steps ($D=500$) and their cosine diffusion schedule in \model. 
We also use their dataset normalization statistics (i.e. normalize the FLAME coefficients by subtracting mean and dividing by standard deviation).
Furthermore, in order to equalize the input conditions, we trained \model with an \emph{extra input condition} - the FLAME identity shape vector $\shapecoeff$, which is also what the original DiffPoseTalk is trained with.
We train a number of models to cover all possible scenarios: 

\paragraph{(1) Models trained without head pose.} 
Consistently to the experiments on THUNDERSET, we find that \model models benefit from the \mts loss, which can be observed in improvement on all lip-sync metrics. 
The improvement is more dramatic in \model-F models, but manifests itself also in the models with trainable audio encoders (\model-F). 
Similarly to our experiments on THUNDERSET, the application of \mts comes with a small reduction in upper-face diversity. 
Finally, \model-T outperforms DiffPoseTalk on all lip-sync metrics. 
The metrics can be seen in \tabref{tab:TFHP}, experiment 1.

\paragraph{(2) Models trained with head pose.}
Since TFHP comes with head pose annotations that correspond to natural head movement during spontaneous speech, we investigate if \model models can also be trained to produce head pose. 
When it comes to lip-sync and face diversity, all findings are consistent with that of the experiment without head pose. 
Additionally, we find that \model is capable of producing head pose with higher beat alignment (BA) and higher diversity (DIV) than DiffPoseTalk. 
This suggests that the \mts loss does not interfere with head pose generation.
See \tabref{tab:TFHP}, experiment 2 for more detail.
Note that the beat alignment and diversity were computed with the global rotation representation converted into the 6D rotation representation. %

\paragraph{(3) Models trained with head pose and speaking style conditioning.}
Finally, we train \model models with the DiffPoseTalk contrastive style features as conditions. 
The results can be found in See \tabref{tab:TFHP}, experiment 3. 
We find that conditioning with the style vector results in better lip-sync performance, lower face diversity, higher beat alignment and lower pose diversity compared to measurements in See \tabref{tab:TFHP}, experiment 2.
All of these are consistent with expectations, since passing in a style specification reduces the distribution of possible outcomes. 
Regardless of the above, the application of \mts has the expected effect - improved lip-sync metrics and lower diversity. 

\paragraph{(4) Co-supervising with a perceptual loss. }

A recent paper by Chae et al.~\cite{chae2025perceptually} proposes a new  self-supervised speech-mesh representation, in form of a bi-modal masked autoencoder. 
The architecture consists of two encoders and two decoders (one for the mesh modality and one for the audio modality). 
The model is trained with two loss terms - the autoencoder reconstruction losses for both modalities and the InfoNCE loss which brings the two modalities into one coherent features space. 
This is similar to what CLIP \cite{Radford2021clip} does with images and text. 
The authors show that their representation can be used in training other speech-driven animation methods. 
They use their frozen autoencoder in training, passing the input audio and output geometry in, extracting the features for both. 
Then, they apply the same InfoNCE loss between the two modalities and use it to co-supervise the talking head avatar system.
In this experiment, we incorporate this perceptual loss into the \model training.
The training loss is then given as: 
\begin{equation}
    \mathcal{L}_{total} = \wmts \lossmts +  w_{perc} \mathcal{L}_{perc} + \lossrec  + \lossvel,
\end{equation}
where $ w_{perc} \mathcal{L}_{perc} $ is the perceptual term. We follow the authors and set the weight $w_{perc} = 0.1 $, a factor of 7 orders of magnitude lower than the vertex loss, which puts the computed losses on a similar scale and hence results in optimal effect (i.e not too strong or too weak). 
For completeness, we also report the new metrics proposed by Chae et al.~\cite{chae2025perceptually} - mean temporal misalignment (MTM) and Perceptual Lip Readability Score (PLRS) along with the other metrics used in this paper. 
MTM measures the temporal misalignment using a Derivative Dynamic Time Warping (DDTW). 
SLRS is the same InfoNCE-based perceptual loss used for training, but computed by a different instance of the same architecture.
The authors released two instances - one to be used as the perceptual loss and one to be used for evaluation with SLRS.
We utilize both of the speech-mesh autoencoder models released by the authors accordingly. 
\tabref{tab:TFHP}, experiment 4 reports the results of this experiment. 
In this experiment we investigate the following: 

\emph{(a) Does training \model with mesh-to-speech result in improvement on metrics proposed by Chae et al.~?} 
Yes, all \model in all settings (across experiments 1-4) improve \model without mesh-to-speech and DiffPoseTalk on both MTM and PLRS. 
The improvement on MTM is expected, since MTM leverages DDTW, which is similar to the DTW metric \cite{bala2022imitator} we use. 
Furthermore, improvement on PLRS indicates that the mesh-to-speech loss results in a perceptual improvement in quality.

\emph{(b)Does training our diffusion-based speech-driven animation method with this perceptual loss improve performance?}
Yes, \model models trained without mesh-to-speech but \emph{with} the perceptual loss improve considerably on PLRS. 
This is expected since PLRS is computed by a model with the same architecture as the network which provides the perceptual loss function. 
We also observe improvement on MTM. 
Furthermore, we observe improvement on our lip-sync related metrics in models with the frozen backbone (-F). 
Models with trainable backbones (-T) do not seem to improve on LVE, CCC, PCC or DTW when the perceptual loss is applied.

\emph{(c) Can \model be trained with both mesh-to-speech and the perceptual loss.}
Yes. Both losses can be applied together. 
The mesh-to-speech loss has a strong impact on LVE, CCC, PCC, DTW and MTM, while the perceptual loss strongly affects PLRS, MTM and with -F models also improves CCC and PCC. 
THUNDER-F with both mesh-to-speech and the perceptual loss appears to get the best of both worlds, 
while \model-T with mesh-to-speech only is competitive with \model-T with both losses.
In any case, applying the mesh-to-speech is beneficial.

\subsubsection{Disentanglement} 
\label{sec:disentanglement}
\begin{figure}[b]
    \offinterlineskip
    \centering
    \includegraphics[width=0.2\columnwidth]{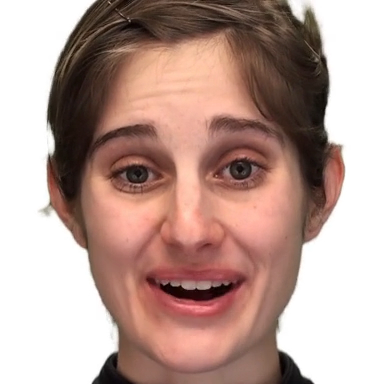}
    \includegraphics[width=0.2\columnwidth]{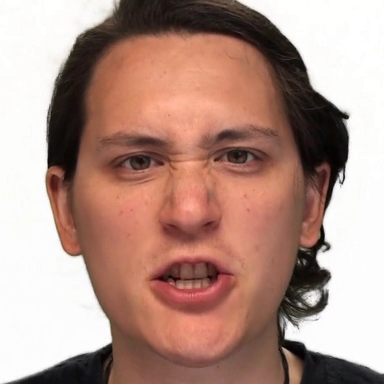}
    \includegraphics[width=0.2\columnwidth]{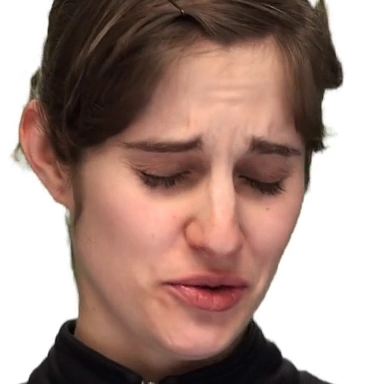}
    \includegraphics[width=0.2\columnwidth]{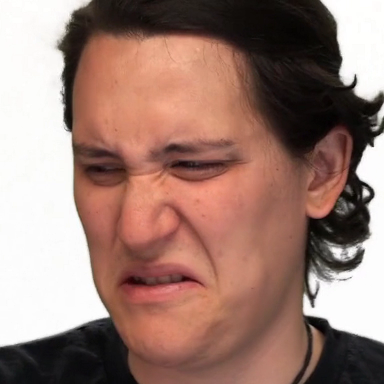}
    \vspace{1em}
    \includegraphics[width=0.2\columnwidth]{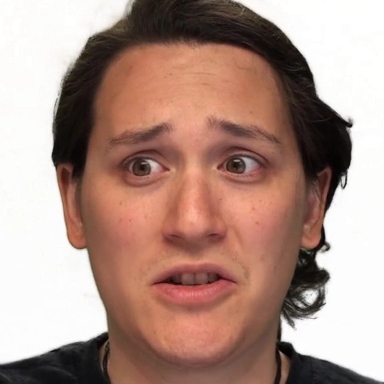}
    \includegraphics[width=0.2\columnwidth]{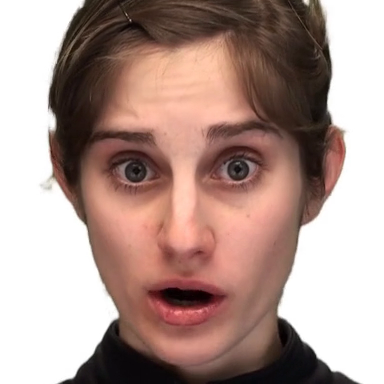}
    \includegraphics[width=0.2\columnwidth]{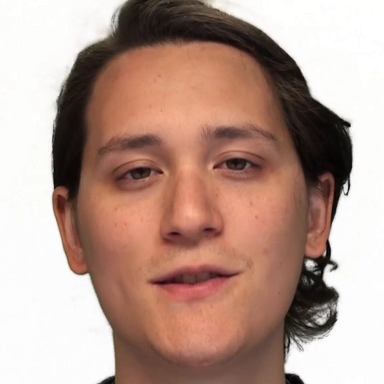}
    \caption{
       \textbf{Media2Face conditioning images}. These images were used as the conditions for the Media2Face* disentanglement experiment in \tabref{tab:disentanglement}. The images were selected out from the RAVDESS test set. Top row from left to right: happy, angry, sad, disgusted. Bottom row: fearful, surprised, calm.
    }
    \label{fig:m2face_condition}
\end{figure}

Ideally, \mts should help preserve high-quality lip-sync in the presence of other editing conditions. In other words, the lip-sync should effectively be disentangled from other conditions. To validate this, we evaluate our re-implementation of Media2Face (THUNDERSET-trained) on the THUNDERSET test set, passing additional conditioning inputs extracted from emotional images from unrelated sequences. 
The conditioning images can be found in \figref{fig:m2face_condition}.
For example results of Media2Face animations, please refer to the supplementary video.
We find that training Media2Face with mesh-to-speech improves lip-sync.
The metrics can be found in \tabref{tab:disentanglement}. 

\begin{table}[t]
\centering
\resizebox{0.49\textwidth}{!}{
\begin{tabular}{|l|ccccc|cc|}
    \toprule
     Name                           & LVE [cm] $\downarrow$ & LIP CCC $\uparrow$ & LIP PCC $\uparrow$ &   DTW [cm] $\downarrow$ &          S-DIV-L [cm] $\downarrow$    & MTM [ms] $\downarrow$ & PLRS $\uparrow$ \\
    \midrule
    M2F-F* orig.                &       \textit{0.611} &     \textit{0.522} &     \textit{0.617} &      \textit{0.221} &               \textit{0.0956}  &    \textit{41.0} &   \textit{0.27}  \\
    M2F-F* w/ m2s orig.      &       \textbf{0.551} &     \textbf{0.558} &     \textbf{0.679} &      \textbf{0.212} &                \textbf{0.027}     &    \textbf{57.5} &  \textbf{0.283} \\
                                   
    \midrule
    M2F-F* angry                &         \textit{1.6} &     \textit{0.307} &     \textit{0.526} &      \textit{0.914} &                \textit{0.151}  &     \textit{88.8} &  \textit{0.173}\\
    M2F-F* w/ m2s angry      &        \textbf{1.46} &     \textbf{0.348} &      \textbf{0.59} &      \textbf{0.741} &                 \textbf{0.08}     &     \textbf{61.7} &  \textbf{0.176}\\
    
    \midrule
    M2F-F* calm                 &        \textit{0.91} &     \textit{0.369} &     \textit{0.537} &      \textbf{0.297} &                \textit{0.103} &    \textit{75.5} &  \textit{0.257} \\
    M2F-F* w/ m2s calm       &        \textbf{0.86} &     \textbf{0.459} &     \textbf{0.638} &      \textit{0.302} &               \textbf{0.0289}    &    \textbf{62.5} &   \textbf{0.27} \\
    
    \midrule
    M2F-F* disgust              &       \textit{0.928} &     \textit{0.314} &     \textit{0.527} &      \textit{0.316} &               \textit{0.0675} &    \textit{77.9} &  \textit{0.249} \\
    M2F-F* w/ m2s disgust    &       \textbf{0.874} &     \textbf{0.426} &     \textbf{0.627} &      \textbf{0.305} &                \textbf{0.019}    &    \textbf{66.6} &  \textbf{0.296} \\
    
    \midrule
    M2F-F* fearful              &        \textbf{1.01} &      \textbf{0.38} &     \textit{0.618} &      \textit{0.315} &                \textit{0.101}  &     \textit{60.8} &  \textbf{0.263} \\
    M2F-F* w/ m2s fearful    &        \textit{1.03} &     \textit{0.353} &     \textbf{0.627} &      \textbf{0.303} &               \textbf{0.0295}    &    \textbf{56.4} &   \textit{0.24}\\    
    \midrule
    M2F-F* happy                &        \textit{1.31} &     \textit{0.201} &     \textit{0.394} &      \textit{0.484} &                \textit{0.152} &    \textit{82.0} &  \textit{0.209}  \\
    M2F-F* w/ m2s happy      &        \textbf{1.11} &       \textbf{0.3} &     \textbf{0.597} &      \textbf{0.269} &                \textbf{0.047}    &    \textbf{66.6} &  \textbf{0.265} \\
    
    \midrule
    M2F-F* sad                  &        \textit{0.96} &     \textit{0.308} &     \textit{0.531} &      \textit{0.363} &                \textit{0.108}  &    \textit{77.3} &  \textit{0.257} \\
    M2F-F* w/ m2s sad        &       \textbf{0.815} &     \textbf{0.428} &     \textbf{0.609} &      \textbf{0.282} &               \textbf{0.0371}     &    \textbf{56.9} &  \textbf{0.287}\\

    \midrule
    M2F-F* surprised            &        \textbf{1.04} &     \textbf{0.391} &     \textit{0.576} &      \textit{0.494} &               \textit{0.0759} &    \textit{63.5} &   \textbf{0.26} \\
    M2F-F* w/ m2s surprised  &        \textit{1.06} &     \textit{0.366} &     \textbf{0.602} &      \textbf{0.362} &               \textbf{0.0278}  &    \textbf{53.7} &  \textit{0.254}\\
    \bottomrule
\end{tabular}
    
}
\caption{{\bf Disentanglement effect.} We feed different input conditions to our versions of Media2Face* models and report the lip-sync metrics for both. \textit{Orig.} denotes an image condition extracted from the video that corresponds to the audio. All other rows are results for denoising with other conditions that come from 7 different emotional images.  Media2Face* with \model exhibits superior lip-sync in most cases. 
Please note that since there is not Ground Truth available for any other setting except for passing the original image (ie. M2F orig.), the metrics are still computed against the original GT. 
That said, since PLRS is a perceptual score, it does not require GT.
However, the consistent improvement in the lip metrics across all condition is still indicative of superior lip-sync.
}
\label{tab:disentanglement}
\end{table}

\subsubsection{Extended Ablation and Sensitivity Analysis}

\begin{table*}[b]
\centering
\resizebox{0.99\textwidth}{!}{
\begin{tabular}{c|l|cccc|cc|cc|cc}
\toprule
  \multicolumn{2}{c|}{ } & \multicolumn{4}{c}{Lip-Sync} & \multicolumn{2}{|c}{Upper-face diversity}  & \multicolumn{2}{|c}{Lip diversity} & \multicolumn{2}{|c}{Face Dynamic Dev.} \\
  \midrule
  Experiment &      Name             &  LVE $\downarrow$   & L-CCC $\uparrow$        & L-PCC $\uparrow$   &  DTW $\downarrow$  & S-DIV-U $\uparrow$  & T-DIV-U $\uparrow$                 & S-DIV-L $\downarrow$       & T-DIV-L $\downarrow$       &  FDD-U $\downarrow$         &  FDD-L $\downarrow$        \\
\midrule

\multirow{4}{*}{\shortstack{(1) THUNDER and \\ mesh-to-speech \\ input space }} 

& THUNDER w/o m2s                    &                   0.879 &              0.359 &              0.568 &               0.329 &             \textbf{0.0419} &            \textbf{0.044} &                        0.21 &                   0.254 &             \textit{0.0118} &                    0.0932  \\
& THUNDER w/ face2s                  &          \textit{0.804} &     \textit{0.411} &     \textit{0.633} &      \textbf{0.285} &                      0.0297 &           \textit{0.0409} &              \textbf{0.128} &          \textit{0.237} &             \textbf{0.0117} &           \textit{0.0827}  \\
& THUNDER w/ exp2s                   &                 0.83    &              0.362 &               0.63 &               0.296 &             \textit{0.0404} &                    0.0398 &                       0.176 &          \textbf{0.228} &                      0.0125 &                    0.0939  \\
& \textbf{THUNDER w/ mouth2s}        &      \textbf{0.802}     &     \textbf{0.426} &     \textbf{0.639} &       \textit{0.29} &                      0.0322 &                      0.04 &              \textit{0.134} &                   0.241 &                      0.0122 &           \textbf{0.0806}  \\
\midrule

\multirow{6}{*}{\shortstack{(2) THUNDER-F \\ mesh-to-speech \\ sensitivity analysis }} 

& THUNDER-F w/o m2s                  &                   0.879 &              0.359 &              0.568 &               0.329 &             \textit{0.0419} &            \textit{0.044} &                        0.21 &                  0.254 &             \textbf{0.0118} &           \textit{0.0932}  \\
& THUNDER-F $ \wmts = 0.1$           &                  0.887  &     \textit{0.375} &     \textit{0.599} &               0.309 &             \textbf{0.0422} &           \textbf{0.0445} &                       0.211 &                  0.258 &                      0.0128 &                    0.0936  \\
& THUNDER-F $ \wmts = 0.5 $          &        \textit{0.805}  &              0.342 &              0.574 &      \textit{0.294} &                      0.0387 &                    0.0383 &                       0.164 &          \textit{0.225} &                      0.0131 &                      0.103  \\
& \textbf{THUNDER-F  $ \wmts= 1.0$}  &        \textbf{0.802}  &     \textbf{0.426} &     \textbf{0.639} &       \textbf{0.29} &                      0.0322 &                      0.04 &              \textbf{0.134} &                   0.241 &             \textit{0.0122} &            \textbf{0.0806}  \\
& THUNDER-F $ \wmts=5.0$             &                  0.858 &               0.32 &              0.572 &               0.313 &                      0.0257 &                    0.0364 &                       0.159 &                    0.23 &                      0.0143 &                      0.105  \\
& THUNDER-F $ \wmts=10.0 $           &                  0.886 &              0.311 &              0.585 &                0.31 &                       0.021 &                     0.035 &              \textit{0.146} &          \textbf{0.219} &                      0.0155 &                      0.105 \\
\midrule 
\multirow{6}{*}{\shortstack{(3) THUNDER-T \\ mesh-to-speech \\ sensitivity analysis }} 
& THUNDER-T w/o m2s                  &                  0.723 &              0.428 &              0.623 &               0.266 &                        0.02 &           \textit{0.0411} &             \textit{0.0656} &                   0.216 &                      0.0118 &                    0.0811  \\
& THUNDER-T $ \wmts= 0.1 $           &                   0.79 &              0.417 &               0.63 &      \textbf{0.252} &             \textbf{0.0279} &           \textbf{0.0431} &                      0.0964 &                   0.221 &             \textbf{0.0117} &           \textbf{0.0729}  \\
& THUNDER-T $ \wmts = 0.5$           &        \textbf{0.693}  &     \textit{0.435} &     \textit{0.641} &               0.263 &             \textit{0.0221} &                    0.0391 &                      0.0662 &          \textit{0.201} &                      0.0118 &           \textit{0.0782}  \\
& \textbf{THUNDER-T $ \wmts=1.0$}    &         \textit{0.709} &     \textbf{0.445} &      \textbf{0.66} &      \textit{0.256} &                       0.021 &                     0.039 &                      0.0669 &                   0.202 &             \textit{0.0118} &                    0.0788  \\
& THUNDER-T $ \wmts=5.0$             &                  0.844 &              0.343 &              0.614 &               0.262 &                      0.0213 &                    0.0351 &                      0.0997 &                   0.203 &                      0.0139 &                    0.0908  \\
& THUNDER-T $ \wmts=10.0$            &                  0.737 &              0.408 &              0.621 &               0.301 &                      0.0207 &                    0.0344 &             \textbf{0.0486} &          \textbf{0.199} &                      0.0147 &                     0.089  \\
\bottomrule
\end{tabular}
}
\caption{{\bf Ablation study and sensitivity analysis.} Here we analyze the effect of the mesh-to-speech loss on the talking head avatar training. 
This experiment is conducted on THUNDERSET.
The top section compares models supervised with mouth-mesh-to-speech (mouth2s), full-mesh-to-speech (face2s) and flame-to-speech (exp2s). 
While it is already present in the main paper text, we include it here for completeness. 
The mid and bottom sections analyze the effect of weight $ w_{m2s} $, with either frozen (-F) or trainable (-T) Wav2Vec2. 
THUNDER-T models exhibit superior lip-sync performance but it comes at the expense of reduced diversity. Furthermore, for both THUNDER-F and THUNDER-T models, increasing $ w_{m2s} $, while beneficial for lip animation quality, comes at an increasing expense of diversity.
Please refer to the Sup.~Video for qualitative comparison. 
}
\label{tab:supmat_ablation}
\end{table*}

\tabref{tab:supmat_ablation} shows the complete ablation and sensitivity analysis of all \model components on THUNDERSET.
\label{sec:extened_ablation}
Here we give a comprehensive evaluation of our design decisions.
(1) The input space for the mesh-to-speech loss, (2) the weight of the mesh-to-speech loss, and (3) the effect if finetuning of the input audio encoder.

\qheading{(1) On mesh-to-speech input space.}
While all three modalities are comparable, \textit{mouth2s} has scores best on the lip-sync metrics, likely thanks to the localized effect on the mouth. \textit{Exp2s} supervision performs slightly worse on lip-sync but results in higher sample diversity. 
Finally, applying supervision through the whole face to speech (\textit{face2s}), while achieving comparable lip-sync scores, scores worse on diversity, specifically S-DIV-U. 
The metric can be found in \tabref{tab:supmat_ablation}, experiment 1.

\qheading{(2) On mesh-to-speech strength.} Applying mesh-to-speech loss, even weakly, results in improved lip-sync metrics (LVE, CCC, PCC and DTW). The stronger the loss, the more the lip-sync metrics improve. However, increasing weights of the mesh-to-speech loss come at the expense of generation diversity (metrics S-DIV, T-DIV). See  \tabref{tab:supmat_ablation}, experiment 2.

\qheading{(3) On finetuning the audio encoder.}
Nearly all recent methods 
finetune the input audio encoder or else they suffer from considerably impaired lip-sync. 
This results in a certain amount of overfitting to the training subjects' voices, which manifests as lower diversity scores (S-DIV-U) compared to the models with frozen Wav2Vec (\model-F). 
In essence, the model's with trainable audio encoders become less stochastic and more deterministic.
Remarkably, the application of the mesh-to-speech loss alleviates the necessity for finetuning Wav2Vec, producing significantly better lip-sync even without finetuning Wav2Vec (\model-T), dramatically reducing the number of training parameters. Applying mesh-to-speech along with finetuning Wav2Vec results in further improvement of lip-sync metrics. 
Refer to \tabref{tab:supmat_ablation}, experiment 3.

Please refer to the supplementary video for visual examples of all the phenomena described above.

\subsubsection{Perceptual Study}

\begin{figure}[!htbp]
    \begin{minipage}{0.99\columnwidth}
    \offinterlineskip
    \centering
    \includegraphics[trim=0.0cm 2.2cm 0.0cm 1.5cm, clip, width=1.\columnwidth]{figs/perceptual/v3/w2vf_reb_no_emote_dfp/likert_lip_sync.pdf}
    \caption{
       \textbf{THUNDER-F perceptual study - lip-sync.} The results of \model-F against other methods. We compare \model-F against methods with both trainable audio encoder
      (FlameFormer-T, THUNDER-T w/o mesh-to-speech) and frozen encoders (FlameFormer-F, THUNDER-F w/o mesh-to-speech). \model-F outperforms 
       other methods on lip-sync, including those with the trainable audio-encoder (denoted -T).  \
    }
    \label{fig:perceptual_thunderF_lip}
    \end{minipage}

\begin{minipage}{0.99\columnwidth}
    \offinterlineskip
    \centering
    \includegraphics[trim=0.0cm 2.2cm 0.0cm 1.5cm, clip, width=1.\columnwidth]{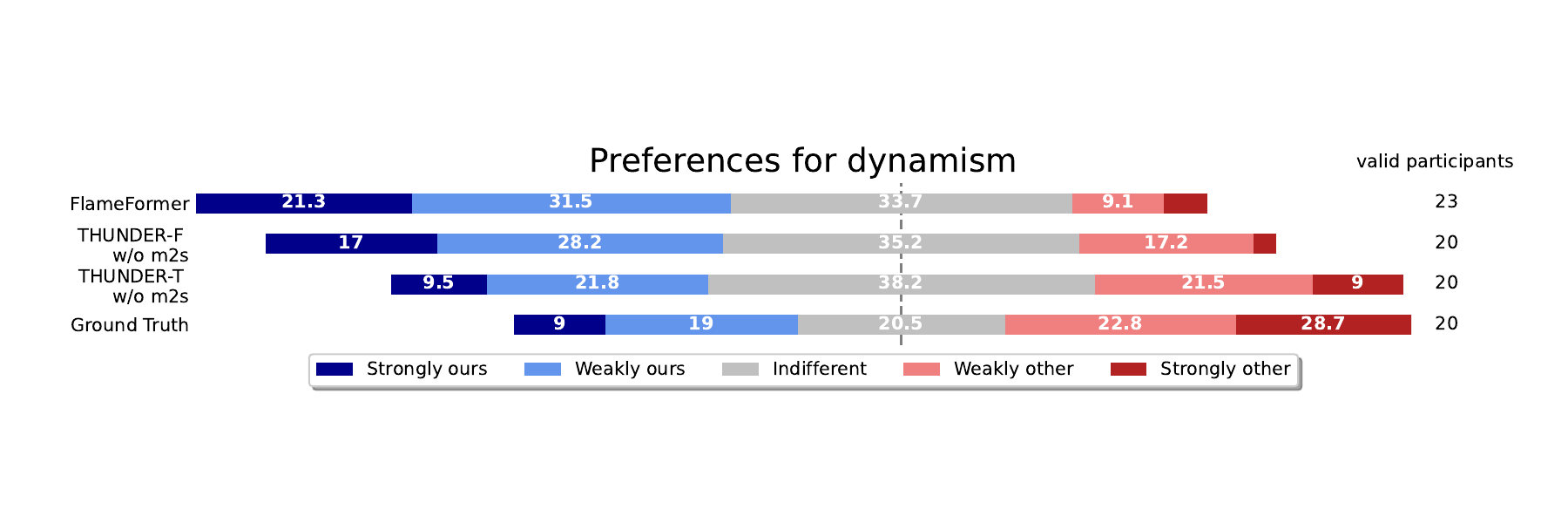}
    \caption{
       \textbf{THUNDER-F perceptual study - dynamism.} 
        The participants found \model-F considerably more dynamic that the deterministic FlameFormer and also more dynamic than \model-F w/o mesh-to-speech. 
        However, ground truth is still rated more dynamic than \model-F. 
    }
    \label{fig:perceptual_thunderF_dyn}
\end{minipage}

\begin{minipage}{0.99\columnwidth}
    \offinterlineskip
    \centering
    \includegraphics[trim=0.0cm 2.2cm 0.0cm 1.5cm, clip, width=1.\columnwidth]{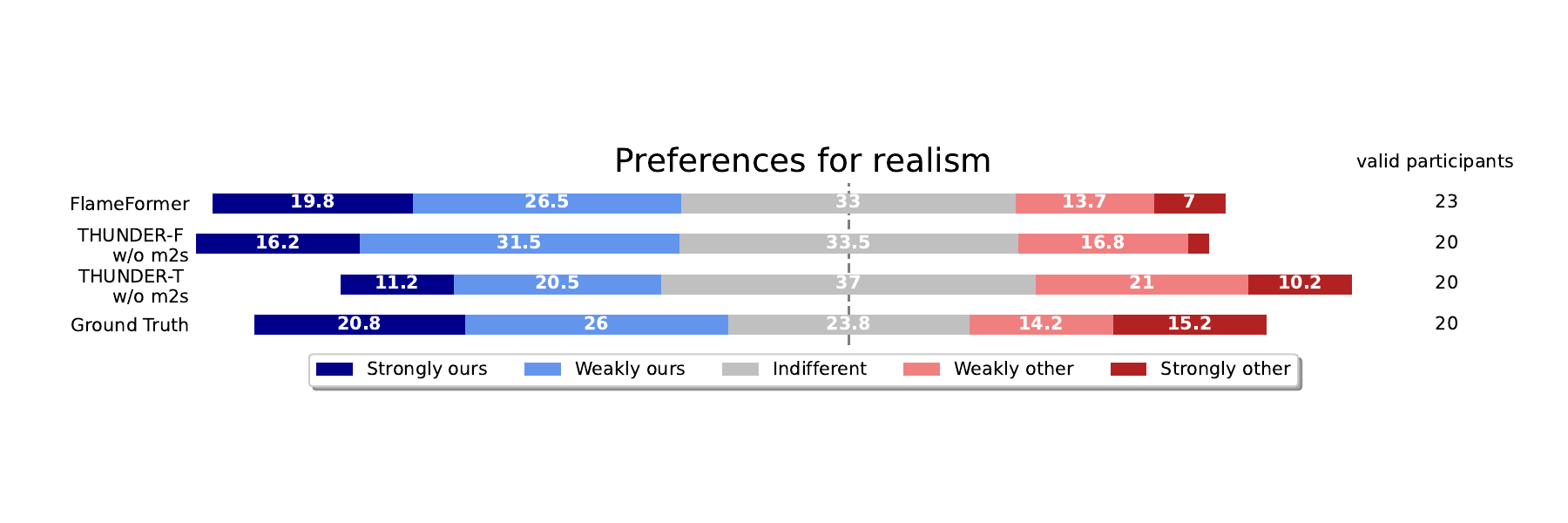}
    \caption{
       \textbf{THUNDER-F perceptual study - realism.} 
        The participants found \model-F considerably more dynamic that the deterministic FlameFormer and also more dynamic than \model-F w/o mesh-to-speech. 
        \model-T w/o mesh-to-speech is rated about as realistic as \model.
        Remarkably, the participants have preferred \model's generation over ground truth in terms of realism, which suggests the training dataset has been exhausted.
    }
    \label{fig:perceptual_thunderF_real}
\end{minipage}
\end{figure}

\begin{figure}[!htbp]
\begin{minipage}{0.99\columnwidth}
    \offinterlineskip
    \centering
    \includegraphics[trim=0.0cm 2.2cm 0.0cm 1.5cm, clip, width=1.\columnwidth]{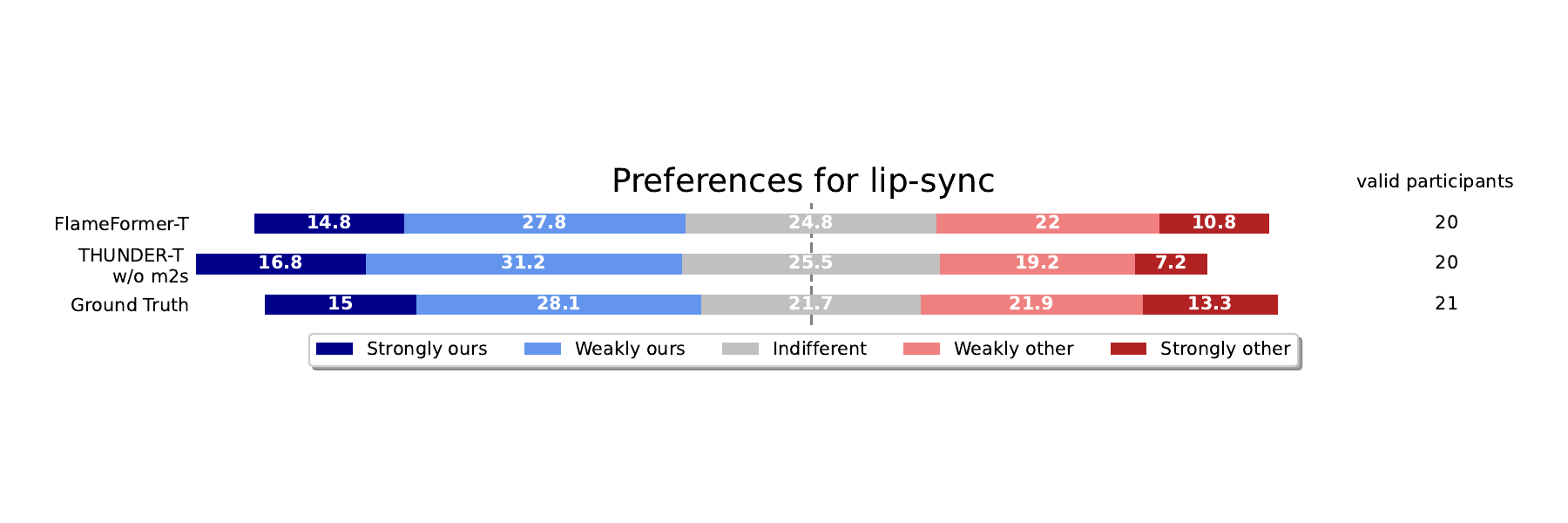}
    \caption{
       \textbf{THUNDER-T perceptual study - lip-sync.} The results of THUNDER-T against other methods. \model-T outperforms its counterpart with mesh-to-speech and the deterministic FlameFormer on lip-sync quality. 
       Additionally, the participants have a slight preference for \model-T over ground truth. 
       This suggests, that THUNDER-T saturates the lip-sync quality of our pseudo-GT.
    }
    \label{fig:perceptual_thunderT_lip}
\end{minipage}

\begin{minipage}{0.99\columnwidth}
    \offinterlineskip
    \centering

    \includegraphics[trim=0.0cm 2.2cm 0.0cm 1.5cm, clip, width=1.\columnwidth]{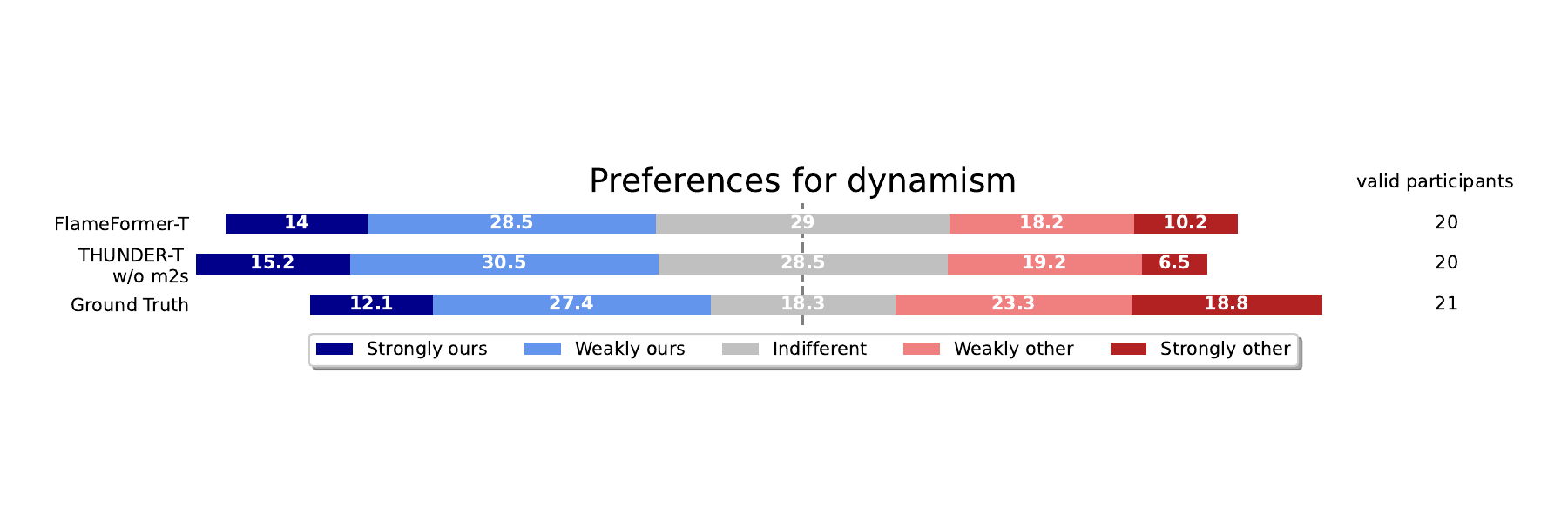}
    \caption{
       \textbf{THUNDER-T perceptual study - dynamism.} The participants rate \model-T to be more dynamic than \model-T without mesh-to-speech and also more dynamic than FlameFormer-T and THUNDER-T w/o mesh-to-speech. 
       Pseudo-GT, however, is preferred over \model-T, suggesting that \model-T's dynamism did not yet exhaust the dynamism of the training dataset.
    }
    \label{fig:perceptual_thunderT_dyn}
\end{minipage}

\begin{minipage}{0.99\columnwidth}
    \offinterlineskip
    \centering
    \includegraphics[trim=0.0cm 2.2cm 0.0cm 1.5cm, clip, width=1.\columnwidth]{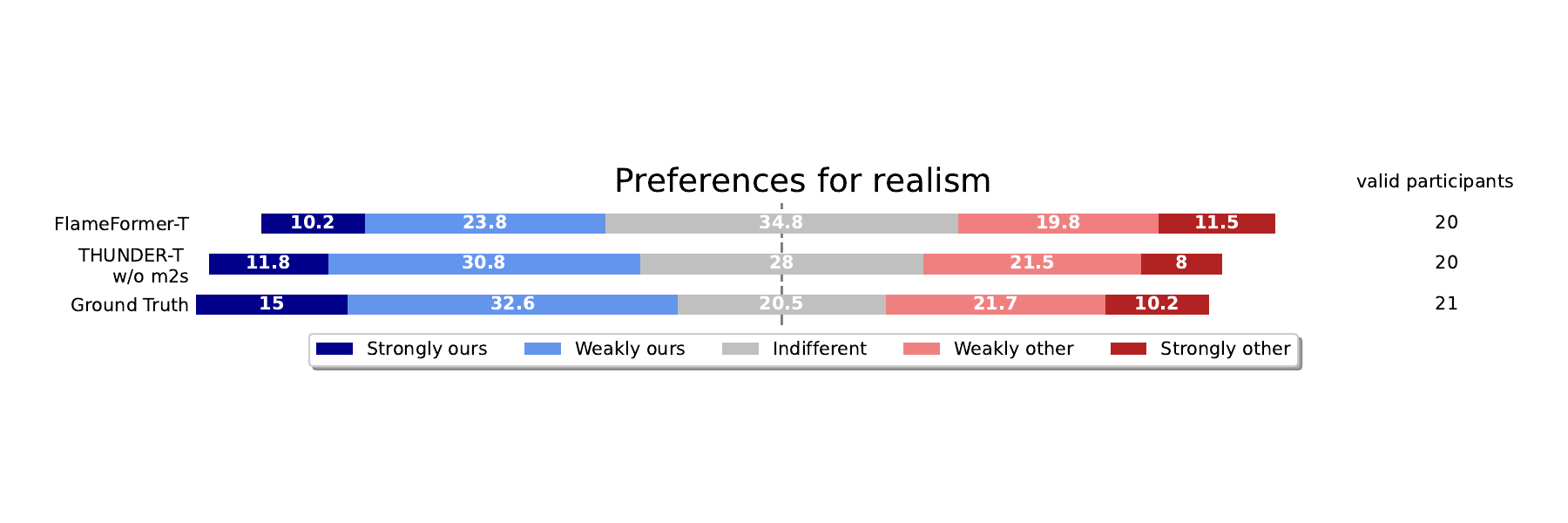}
    \caption{
       \textbf{THUNDER-T perceptual study - realism.} The participants prefer the degree of realism of THUNDER-T's outputs over the deterministic FlameFormer-T and diffusion-based THUNDER without mesh-to-speech.
    }
    \label{fig:perceptual_thunderT_real}
\end{minipage}

\end{figure}

\begin{figure*}[t]
    \offinterlineskip
    \centering
    \includegraphics[width=2.\columnwidth]{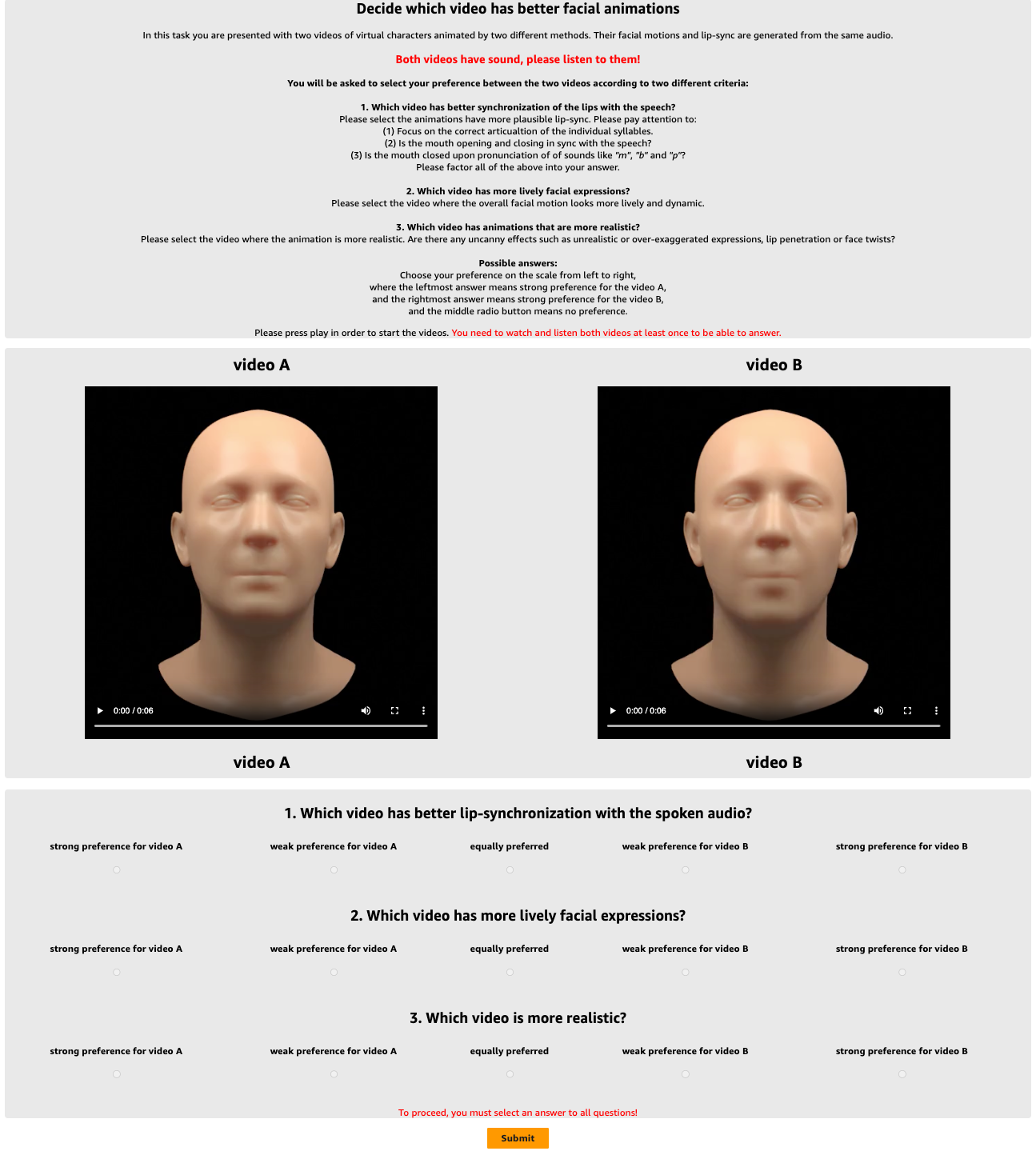}
    \caption{
       \textbf{The perceptual study web template.} 
    }
    \label{fig:perceptual_template}
\end{figure*}

\label{sec:supmat_perceptual}
Here we provide the rest of our comprehensive perceptual evaluation of \model. 
We run a perceptual study on Amazon Mechanical Turk. The participants are shown two videos side-by-side. One corresponding to \model and one to a baseline. 
The left-right order is randomized. 
The participants must finish watching both videos before being allowed to rate the videos on a five-point Likert scale (strong preference for left, weak preference for left, indifference, weak preference for right and strong preference for right). 
We ask the participants to rate three aspects of the animation - lip-sync, dynamism and realism. The exact task description can be seen in the study template in \figref{fig:perceptual_template}.
The participants are shown 20 comparisons, generated from test audios, which where randomly selected from the RAVDESS and TCD-TIMIT test sets.
In addition to that, we repeat the first 4 comparisons at the end of the study and discard the first 4 responses.
This gives the participants a few examples such that they can adjust to the task. 
Additionally, we include 4 catch trials, where one animation is selected from Ground Truth and the other animation is clearly wrong.  
We discard the participants that are wrong on more than one catch trial.
The complete results of the perceptual study for \model for lip-sync, dynamism and realism can be found in Figures \ref{fig:perceptual_thunderF_lip}, \ref{fig:perceptual_thunderF_dyn} and \ref{fig:perceptual_thunderF_real} respectively.  
The results for \model-T are given in Figures \ref{fig:perceptual_thunderT_lip}, \ref{fig:perceptual_thunderT_dyn} and \ref{fig:perceptual_thunderT_real}.
The results validate our qualitative and quantiative findings - \model models exhibit better lip-sync, realism and dynamism when compared to baselines.

\subsubsection{Comparison to other methods.}
Comparing different speech-driven animation methods is a difficult task, especially if trained on a different dataset and especially if the data was acquired in different ways (4D scans, pseudo-GT of different reconstruction methods, etc.). 
Having a fair apples-to-apples comparisons stemming from measurements done against the (pseudo-) ground truth is therefore difficult. 
Regardless, for the sake of completeness, we compare \model to the official releases of recent methods that were also trained  on pseudo-GT, namely EMOTE \cite{danecek2023emote} (a deterministic SOTA) and DiffPoseTalk \cite{sun2024diffposetalk} (stochastic SOTA). 
EMOTE is a FaceFormer-inspired architecture with additional input conditions for emotions and intensity, trained with a content-emotion disentanglement mechanism. It was trained on pseudo-GT reconstructions of MEAD. 
The pGT was acquired with the same INFERNO tracker, making the resulting metrics comparable.
The macro-architecture of DiffPoseTalk's diffusion is very similar to \model and apart from the additional style conditioning input, it differs only in small details (such as conditioning with the FLAME shape vector, noise schedule, number of denoising steps, etc.).
The most important difference is that it was trained on a different dataset, and the reconstructions were produced by a combination of SPECTRE \cite{filntisis2022visual} and MICA \cite{MICA:ECCV2022} and hence look qualitatively very different. 
As such, the DiffPoseTalk metrics likely dominated by the qualitative difference, and are only listed for completeness. 
Please note that for this experiment, DiffPoseTalk was run without the style condition (to match the setting of \model).
Finally, we include a FlameFormer trained on our data, and \model trained without \mts.  
The results are reported in Tab.~\ref{tab:talkinghead}.

\begin{table*}[t] %
\centering
\resizebox{0.99\textwidth}{!}{
\begin{tabular}{l|c|c|cccc|cc|cc|ccc}

\specialrule{.15em}{.1em}{.1em}
  \multicolumn{3}{c|}{ } & \multicolumn{4}{c}{Lip-Sync} & \multicolumn{2}{|c}{Sample diversity}  & \multicolumn{2}{|c}{Temporal diversity} & \multicolumn{2}{|c}{Face Dynamic Dev.} \\

\specialrule{.15em}{.1em}{.1em}
        Name         & Dataset    & Input    &   LVE $\downarrow$  & L-CCC $\uparrow$    & L-PCC $\uparrow$   &  DTW $\downarrow$   & S-DIV-U $\uparrow$         & T-DIV-U $\uparrow$        & S-DIV-L $\downarrow$        & T-DIV-L $\uparrow$       &  FDD-U $\downarrow$           &  FDD-L $\downarrow$        \\

\specialrule{.15em}{.1em}{.1em}
        FlameFormer* & THUNDERSET & \shortstack{audio, \\ one-hot speaker ID}  & \textit{0.809} &     \textit{0.368} &      \textit{0.57} &      \textit{0.291} &                      0.0271 &                    0.0372 &              \textbf{0.132} &                   0.239 &              \textbf{0.0117} &           \textit{0.0909}  \\
\midrule
               EMOTE & MEAD &  \shortstack{audio, \\ one-hot speaker ID, \\ emotion and intensity} &          1.06  &              0.221 &              0.442 &               0.466 &             \textbf{0.0563} &                    0.0401 &                       0.248 &                    0.24 &                      0.0141 &                     0.104  \\
\midrule
        DiffPoseTalk & TFHP &  \shortstack{audio, shape vector} &            1.15 &              0.255 &               0.38 &               0.316 &                      0.0345 &           \textbf{0.0479} &                       0.226 &          \textbf{0.308} &                    0.0141   &                     0.102 \\
\midrule 
 THUNDER-F w/o m2s    & THUNDERSET &  audio only &     0.879 &              0.359 &              0.568 &               0.329 &             \textit{0.0419} &            \textit{0.044} &                        0.21 &          \textit{0.254} &             \textit{0.0118} &                    0.0932  \\
\midrule
          THUNDER-F  & THUNDERSET &   audio only& \textbf{0.802} &     \textbf{0.426} &     \textbf{0.639} &       \textbf{0.29} &                      0.0322 &                      0.04 &              \textit{0.134} &                   0.241 &                      0.0122 &           \textbf{0.0806}  \\

\specialrule{.15em}{.1em}{.1em}
\end{tabular}

}
\caption{{\bf Quantitative comparison of \model with other speech-driven avatar methods.} 
Asterisk* indicates we have re-implemented the baseline and trained it on our dataset. \textit{(1) On lip-sync.} \model outperforms all baselines on lip-sync metrics (LVE, CCC, PCC, DTW). 
\textit{(2) On expression diversity.} DiffPoseTalk exhibits higher temporal face diversity T-DIV-U and L-DIV-U, likely due to the training data reconstructed from videos in-the-wild, which contains more dynamic motions than our in-the-lab reconstructions. 
EMOTE exhibits high sample upper face diversity S-DIV-U and expression diversity S-DIV-EXP which is due to the random sampling of speaking/emotion/intensity styles, which while high, may not be fitting to the particular input audio. 
\model w/o m2s and \model follow next in S-DIV-U.  
DiffPoseTalk (the official model) performs comparatively low on lip-sync and well on both temporal and sample diversity, but the result is likely skewed by the fact that it was trained on different dataset with different types of reconstruction. 
}
\label{tab:talkinghead}
\end{table*}

\subsubsection{Qualitative Results}
\label{sec:qualitative_sup}
In this section we provide additional qualitative results and comparisons. 
Furthermore, we strongly recommend watching the supplemental video for an audible and in-motion qualitative results and comparisons.

\begin{figure*}[t]
    \offinterlineskip
    \centering
    \includegraphics[width=1.99\columnwidth]{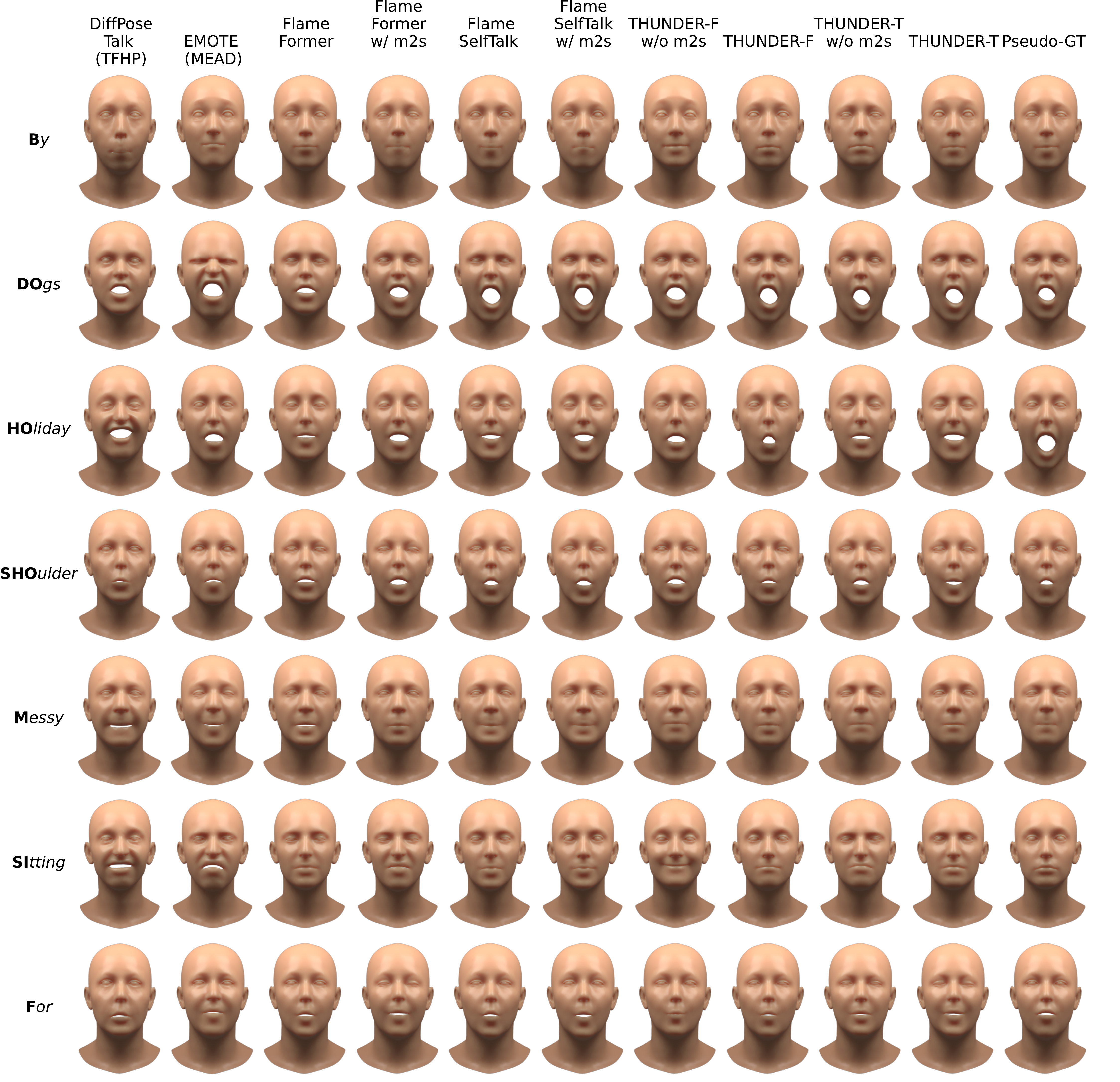}
    \vspace{-0.3cm}
    \caption{
        {\bf Qualitative comparison on THUNDERSET.} This figure shows the comparison between baselines, our model and GT for selected utterances. 
        Note that DiffPoseTalk was trained on TFHP and EMOTE on MEAD. Supplemental PDF and video contain more qualitative comparisons.
    }
    \label{fig:comparison_sup}
\end{figure*}

\qheading{Qualitative Comparison on THUNDERSET.}
\figref{fig:comparison_sup} is a super-set of the figure listed in the main paper (\figref{fig:comparison}).

\begin{figure*}[t]
    \offinterlineskip
    \centering
    \includegraphics[width=2.\columnwidth]{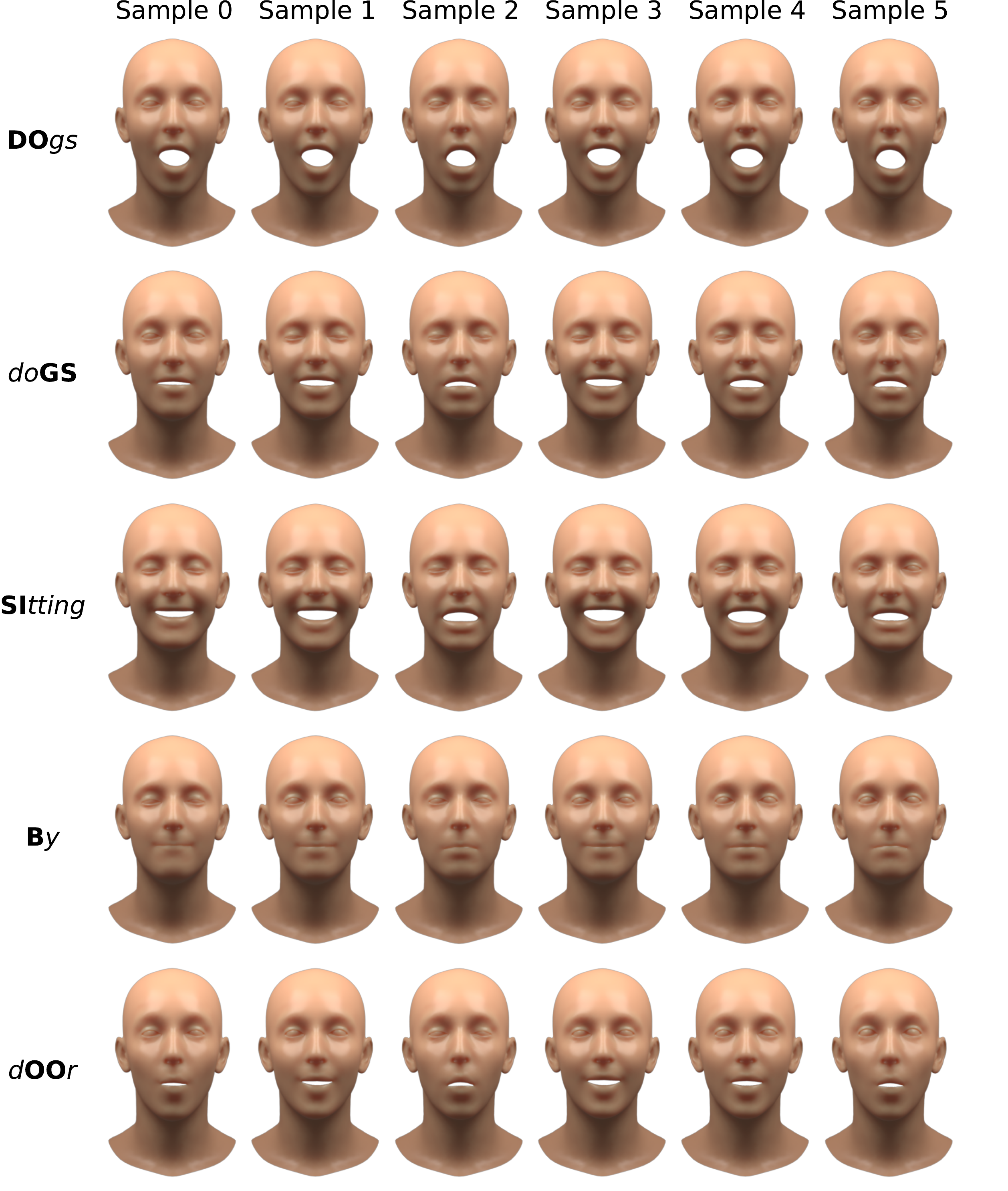}
    \caption{
       \textbf{Diversity of outputs.} \model is capable of generating multiple plausible animations per audio. Each row of this figure contains 6 different generations of the same frame denoised from different initial noise samples (the audio condition remains the same for all). 
    }
    \label{fig:diversity}
\end{figure*}

\qheading{Diversity.}
Input speech may originate from many different expression and even emotions. A well-trained stochastic model should be able to account for that and generate multiple plausible animations, possibly with different facial expressions. \figref{fig:diversity} demonstrates \model's ability to do so.

\begin{figure*}[t]
    \offinterlineskip
    \centering
    \includegraphics[width=2.\columnwidth]{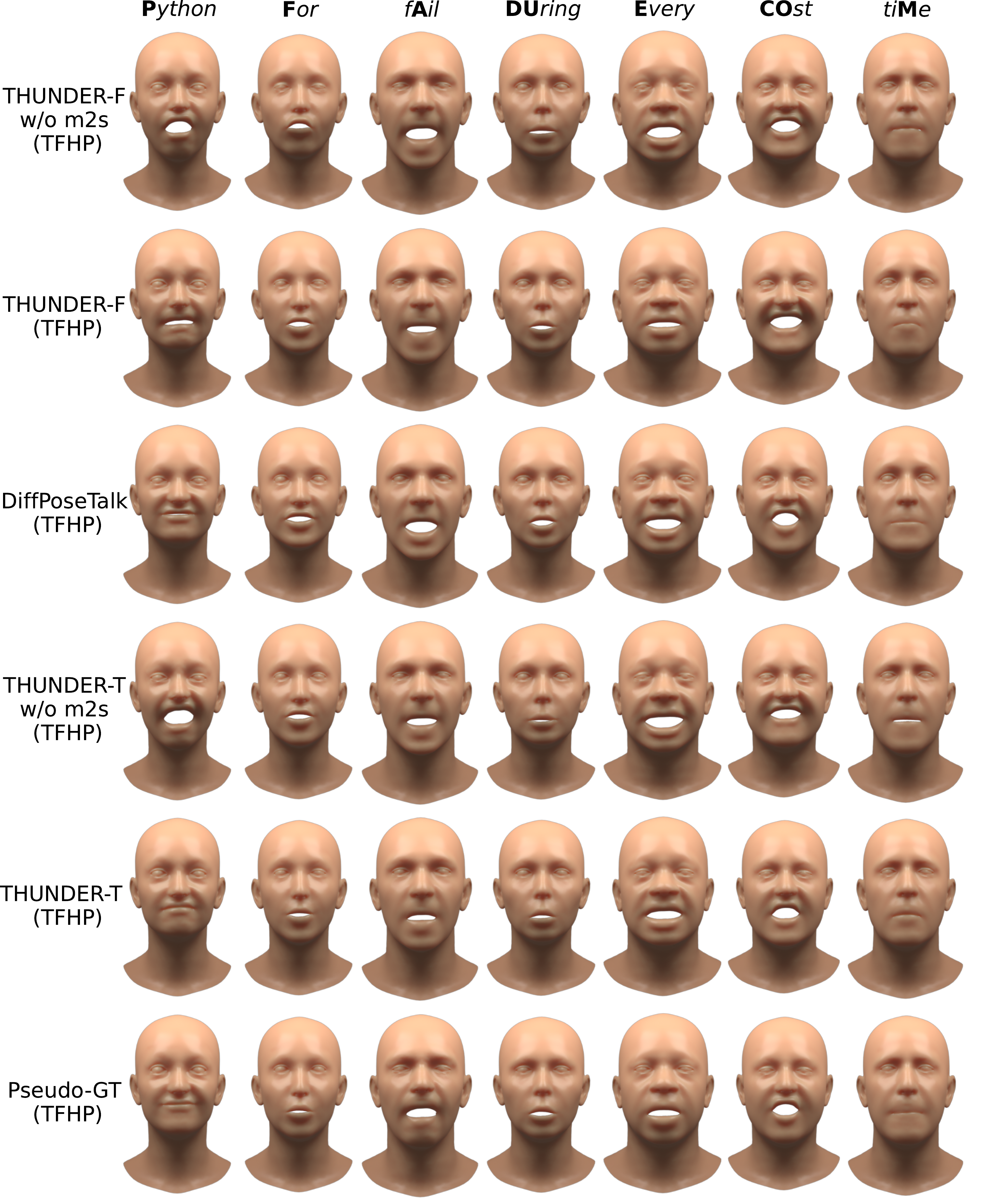}
    \caption{
       \textbf{TFHP models trained without head pose. }
    }
    \label{fig:tfhp_no_pose_models}
\end{figure*}

\begin{figure*}[t]
    \offinterlineskip
    \centering
    \includegraphics[width=2.\columnwidth]{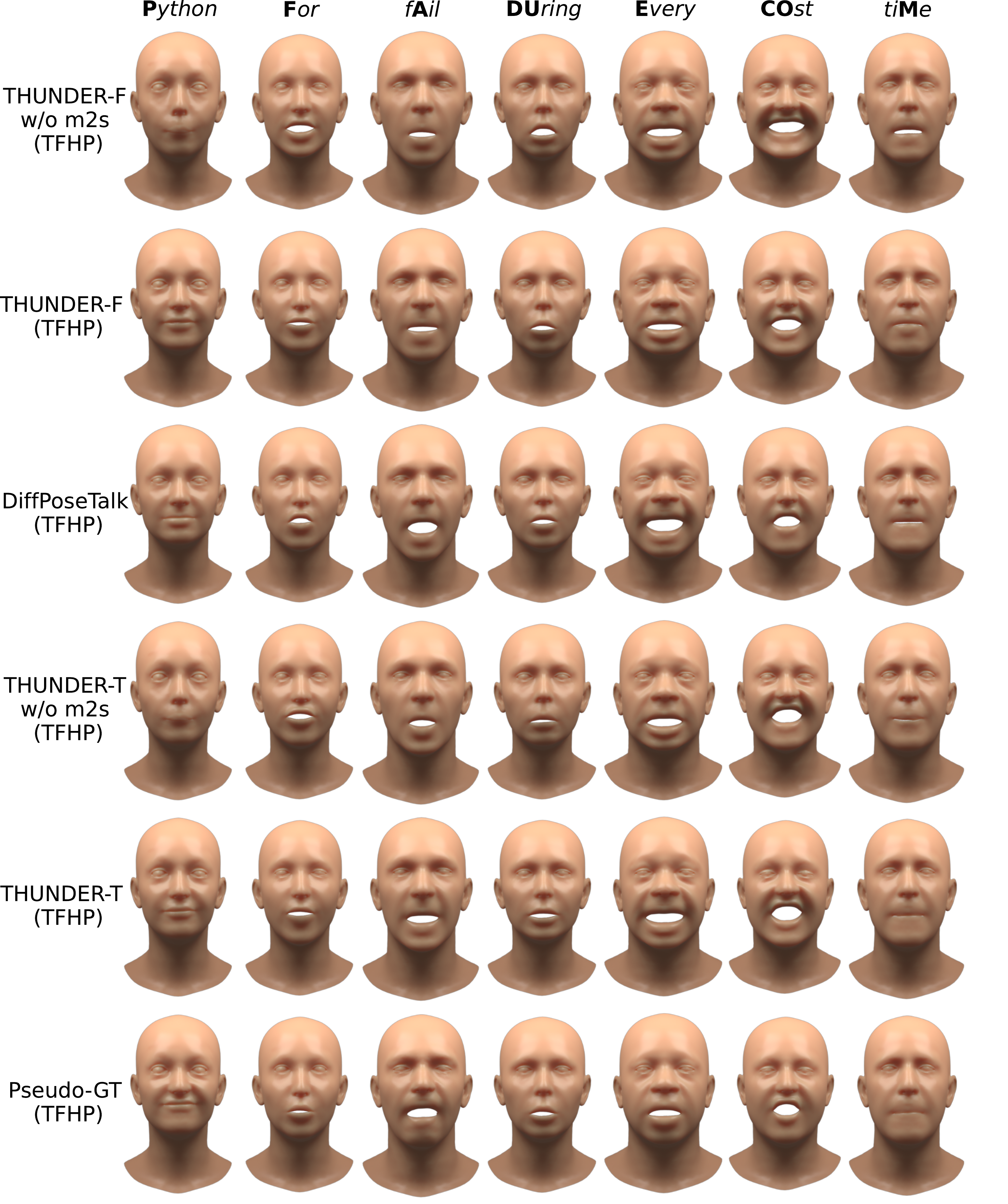}
    \caption{
       \textbf{Results of TFHP-trained models, trained with head pose.} The results are shown in canonical space, i.~e.~ without the generated head pose.
    }
    \label{fig:tfhp_pose_models_unposed}
\end{figure*}

\begin{figure*}[t]
    \offinterlineskip
    \centering
    \includegraphics[width=2.\columnwidth]{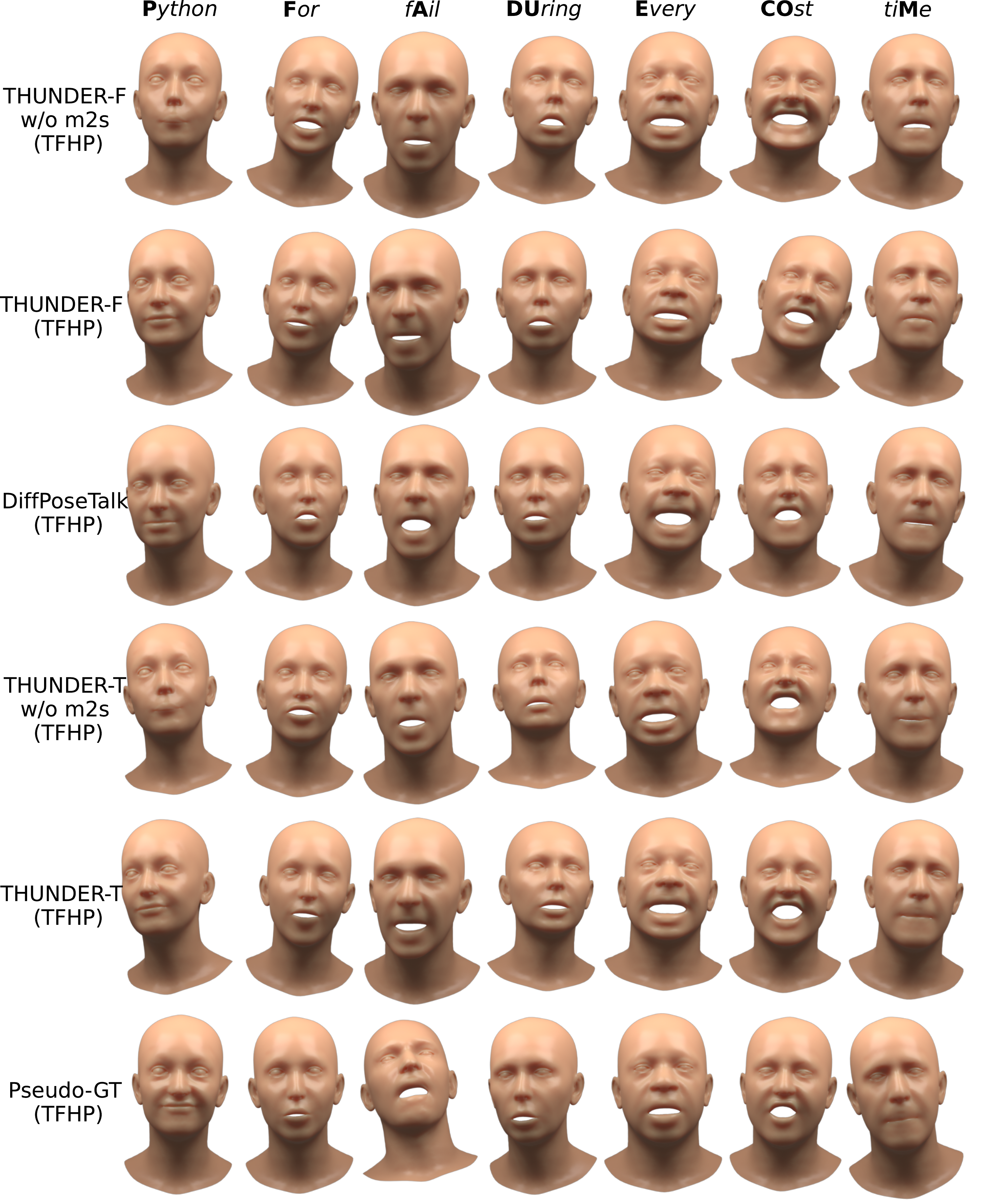}
    \caption{
       \textbf{TFHP models trained with head pose}. The results are shown posed with the predicted pose (or pseudo-GT pose in case of pseudo-GT).
    }
    \label{fig:tfhp_pose_models_posed}
\end{figure*}

\qheading{Qualitative Comparison on TFHP.}
Here we provide qualitative results of models trained on TFHP (the same models from Sec.~\ref{sec:otherdatasets}). 
We select a few utterances from various TFHP test-set sequences. We show results of models trained without head pose in \figref{fig:tfhp_no_pose_models}, the \emph{unposed} results of models trained with pose in   \figref{fig:tfhp_pose_models_unposed}, and finally the same results with the generated pose in \figref{fig:tfhp_pose_models_posed}.

\qheading{Supplementary Video.}
Our video contains exhaustive set of results which demonstrate the benefits of mesh-to-speech and analysis-by-audio-synthesis. The reader is encouraged to watch the video for our qualitative evaluation.

\section{Architecture of Baselines}

\label{sec:baseline_arch}

\begin{figure*}
    \centerline{
    \includegraphics[width=2\columnwidth]{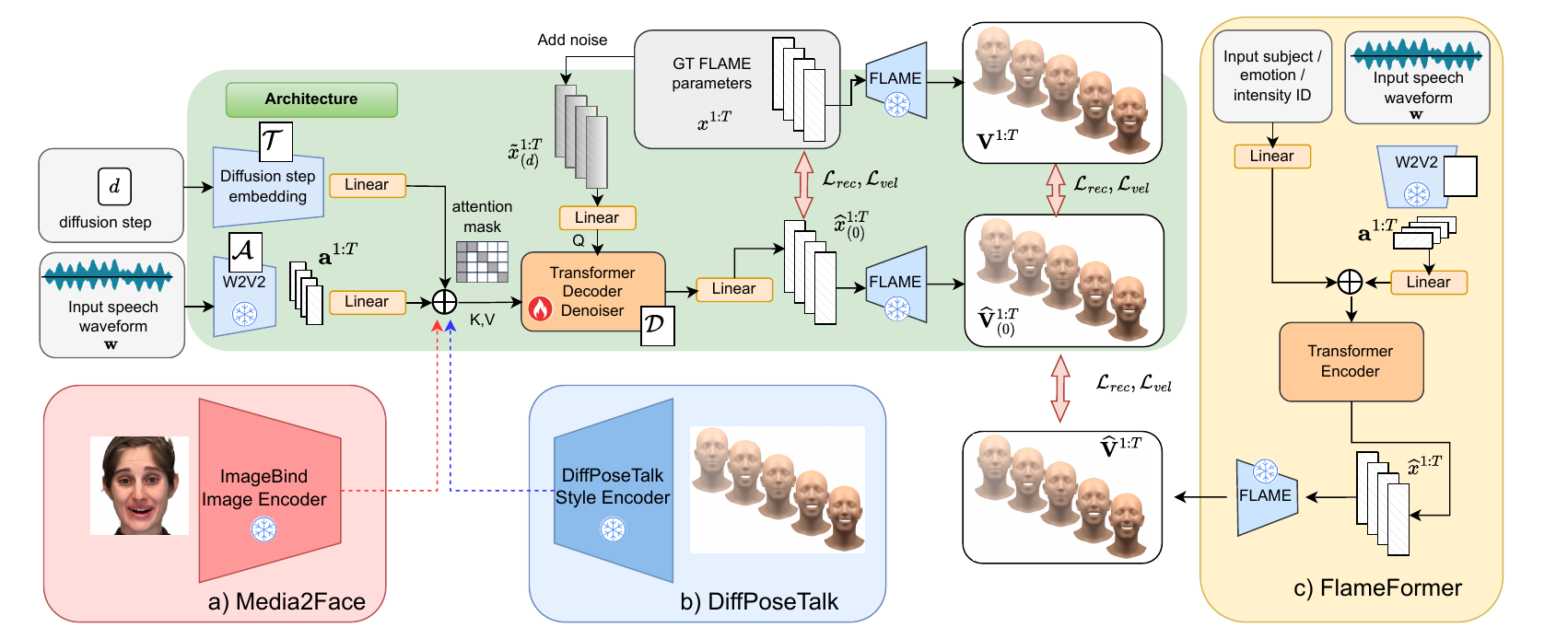}}
    \vspace{-0.1in}
    \caption{
        \textbf{Architecture of re-implemented baselines.} (a) Our version of Media2Face \cite{zhao2024media2face} utilized the same architecture as \model but takes an extra ImabeBind feature on the input. (c) Our re-implementation of DiffPoseTalk \cite{sun2024diffposetalk} is also based on \model and takes an extra style feature from a pretrained style encoder. (c) FlameFormer is an adapted FaceFormer \cite{fan2022faceformer}. FlameFormer predicts FLAME expression instead of full vertex space. Furthermore, FlameFormer adapts a non-autoregressive BERT-like prediction mechanism (akin to EMOTE \cite{danecek2023emote}).
    }
    \label{fig:other_methods_arch}
\end{figure*}

\figref{fig:other_methods_arch} shows the architecture of the baselines we reimplemented in this paper.  

\qheading{Media2Face.} 
Our reimplementation of Media2Face is built on our diffusion architecture. In training, we provide additional input of ImageBind \cite{imagebind_Girdhar2023} features extracted from images of the corresponding videos of the training set. Like the audio condition, this condition is dropped 20\% of the time.

\qheading{DiffPoseTalk.} 
Our reimplementation of DiffPoseTalk is also built on our diffusion architecture. It is trained in two stages. In the first stage we train the contrastive style encoder as proposed by the authors \cite{sun2024diffposetalk}. In training of DiffPoseTalk, we provide the additional style feature on the input. The style feature is extracted from the GT animation sequence. Like the audio condition, the style condition is dropped 20\% of the time.

\qheading{FlameFormer.} 
Our FlameFormer architecture is based on FaceFormer. Like FaceFormer, FlameFormer is a transformer-based deterministic speech-driven animation network that takes addition categorical style condition on the input. Instead of predicting the full vertex space like the original FaceFormer does, FlameFormer predicts FLAME 3DMM coefficients $\expparam$. Additionally, instead of using the autoregressive loop with a transformer decoder, FlameFormer's decoder is a transformer encoder, which eliminates the need for the autoregressive formulation. This does not degrade performance and makes the model more efficient, as was already shown by Daněček et al.~\cite{danecek2023emote}, which have proposed the FlameFormer baseline in their paper. Furthermore, we extend FlameFormer's conditioning inputs - the subject identity (one-hot vector) is complemented with a one-hot vector for 8 basic emotions and three intensities.

\section{Data acquisition} 
\label{sec:data_acquisition}
In this section we describe the rationale behind our choice of methodology to reconstruct the 3D faces from videos.

\subsection{Choice of methodology}
Acquiring high quality 4D scans of sufficient scale and richness is costly, time-consuming and requires specialized hardware. 
Hence, recent speech-driven animation works \cite{sun2024diffposetalk, danecek2023emote, zhao2024media2face, yang2024probabilistic} have turned to pseudo-GT recovered from videos.
This choice comes with an important decision - which face reconstruction methodology to use. 
There are several options but two of the most viable options are: 

\textit{1) Optimization-based 3DDM fitting.} Approaches like these (for instance fitting with analysis-by-image-synthesis) have been proposed more than two decades ago \cite{VetterBlanz1998, Blanz2002, Blanz2003}. While these options are viable and would produce good results, the optimization-based process is rather inefficient when deployed to many hours of video. This remains true even for the contemporary optimization-based trackers, such as the MICA Metrical Tracker \cite{MICA:ECCV2022} or Pixel3DMM \cite{giebenhain2025pixel3dmm}.

\textit{2) Off-the-shelf face 3DMM regressors.}
Due to the computational intensity of optimization-based fitting and thanks to the recent advance in deep-learning based in-the-wild face reconstruction methods \cite{Sanyal2019_RingNet, Feng2021_DECA, EMOCA:CVPR:2021, MICA:ECCV2022, filntisis2022visual, zhang2023tokenface, Retsinas_2024_CVPR_Smirk, Tewari2017, Tewari2018, Tewari2019_FML, Deng2019}. 
While this is still an active area of research, and no method produces results that could be considered as accurate as 4D scans, 
recent years have brought enough advancement so that they can be employed for speech-driven animation dataset construction.

\subsection{Discussion of existing FLAME regressors}
Since \model utilizes FLAME \cite{FLAME:SiggraphAsia2017}, our discussion will focus on FLAME regressors (trackers) only.  
There are several systems that have been used for talking head avatar research. 
No tracker is perfect. 
Each tracker produces different kinds of errors but despite that, the recent trackers are finally good enough to be used for talking head avatar research. 
In fact, a few previous methods have already made use of them \cite{danecek2023emote, mediapipe, sun2024diffposetalk}.
Here we briefly discuss some of the most applicable candidates:
    
    \qheading{DECA} \cite{Feng2021_DECA} is the first FLAME-based regressor trained with the self-supervised "analysis-by-image-synthesis" loop. While this paper was SOTA at the time of its release, it still had considerable artifacts, especially when it comes to the richness of facial expression and emotions and quality of lip animations. Furthermore, the stability of the identity prediction was not stable when applied to videos.

    \qheading{EMOCA} \cite{EMOCA:CVPR:2021} is a follow-up of DECA, capable of reconstructing rich emotions thanks to its emotion-consistency loss. However, it can exaggarate expressions and the lip-sync is not good enough. EMOCAv2 later addressed the lip-sync issue by incorporating the perceptual lip reading loss from SPECTRE, obtaining the best of both worlds. However, similarly to DECA and SPECTRE, EMOCAv2 exhibits unstable identity prediction over the course of a video.

    \qheading{SPECTRE} \cite{filntisis2022visual} applies a paradigm similar to EMOCA but leverages a lip-reading consistency loss instead. It produces good lip-readable animations, but unfortunately often exaggerates and over-articulates, is not capable of reconstructing rich emotional expressions and the disentanglement between expression and identity is rather poor, which often makes the predicted expression vector compensate the inaccuracies in identity predictions.

    \qheading{EMICA} \cite{inferno2023} is a combination of several methods' contributions. It combines the benefits of DECA, EMOCA and SPECTRE to achieve high quality emotions and lip-sync. Additionally, EMICA makes use of MICA \cite{MICA:ECCV2022} for a consistent identity prediction. The system is trained jointly and publicly available. 

    \qheading{SMIRK} \cite{Retsinas_2024_CVPR_Smirk} SMIRK is a recent reconstruction method which uses a neural renderer and in-the-loop neural data augmentation to produce high quality reconstructions.

\subsection{Data acquisition in previous works}
As a result of the tremendous progress of in-the-wild face reconstruction, some of the recent SOTA speech-driven animation methods have already turned to pseudo-GT as the primary source of data. For instance:

    \qheading{EMOTE} \cite{danecek2023emote} employed the EMICA model to reconstruct the MEAD \cite{kaisiyuan2020mead} dataset.

    \qheading{Yang et al.}~\cite{yang2024probabilistic} utilized both DECA and SPECTRE to reconstruct LRS3 \cite{afouras2018lrs3}.
    
    \qheading{DiffPoseTalk} \cite{sun2024diffposetalk} employed MICA and SPECTRE separately and then combined the shape predictions of MICA and expression predictions of SPECTRE to reconstruct the TFHP dataset.

We opt for EMICA as it provides a good compromise - rich emotions, good lip animations, consistent shape identity and temporal consistency of the identity prediction. 
While not artifact-free, EMICA is a good option for building a large-scale dataset for 3D talking head avatar research.

\end{document}